\documentclass[usenatbib]{emulateapj}
\slugcomment{{\sc Accepted by ApJ:} 2011 November 26} 
\providecommand{\comment}[1]{{\bf(#1)}}
\renewcommand{\comment}[1]{}
\usepackage{graphicx}
\usepackage{aas_macros}
\usepackage{amsmath}
\usepackage{amssymb}
 
\newcommand{\chandra}{\textit{Chandra}}
\newcommand{\xmm}{\textit{XMM-Newton}}
\newcommand{\ergscm}{$\mathrm{erg\;s^{-1} cm^{-2}}$}
\newcommand{\cmsq}{cm$^{-2}$}
\newcommand{\ergs}{$\mathrm{erg\;s^{-1}}$}

\newcommand{\dd}{\mathrm{d}}
\newcommand{\NH}{N_\mathrm{H}}

\newcommand{\degsq}{deg$^2$}
\newcommand{\msun}{\mathcal{M}_{\sun}}
\newcommand{\Msun}{$\msun$}
\newcommand{\mstel}{\mathcal{M}_*}
\newcommand{\Mstel}{$\mstel$}
\newcommand{\mbh}{\mathcal{M}_\mathrm{bh}}
\newcommand{\Mbh}{$\mbh$}
\newcommand{\kcorrect}{\textsc{kcorrect}}
\newcommand{\fagn}{$f_\mathrm{AGN}$}
\newcommand{\lx}{L_\mathrm{X}}
\newcommand{\Lx}{$L_\mathrm{X}$}
\newcommand{\giv}{\;|\;}
\newcommand{\lamedd}{\lambda_\mathrm{Edd}}
\newcommand{\Lamedd}{$\lamedd$}
\newcommand{\amend}[1]{{#1}} 
\newcommand{\moreamend}[1]{{#1}} 

\shorttitle{PRIMUS: The dependence of AGN accretion on host stellar mass and color}
\shortauthors{Aird et al.}

\begin{document}

\title{PRIMUS: The dependence of AGN accretion on host stellar mass and color}

\author{
James Aird\altaffilmark{1}, 
Alison L. Coil\altaffilmark{1,2}, 
John Moustakas\altaffilmark{1}, 
Michael R. Blanton\altaffilmark{3}, 
Scott M. Burles\altaffilmark{4}, 
Richard J. Cool\altaffilmark{5,6}, 
Daniel J. Eisenstein\altaffilmark{7}, 
M. Stephen M. Smith\altaffilmark{1},
Kenneth C. Wong\altaffilmark{8}, 
Guangtun Zhu\altaffilmark{3}
}
\altaffiltext{1}{Center for Astrophysics and Space Sciences, Department of Physics, University of California, 9500 Gilman Dr., La Jolla, San Diego, CA 92093}
\altaffiltext{2}{Alfred P. Sloan Foundation Fellow}
\altaffiltext{3}{Center for Cosmology and Particle Physics, Department of Physics, New York University, 4 Washington Place, New York, NY 10003}
\altaffiltext{4}{D.E. Shaw \& Co., L.P., 20400 Stevens Creek Blvd., Suite 850, Cupertino, CA 95014}
\altaffiltext{5}{Department of Astrophysical Sciences, Princeton University, Peyton Hall, Princeton, NJ 08544}
\altaffiltext{6}{Hubble Fellow and Princeton-Carnegie Fellow}
\altaffiltext{7}{Harvard College Observatory, 60 Garden St., Cambridge, MA 02138}
\altaffiltext{8}{Steward Observatory, The University of Arizona, 933 N. Cherry Ave., Tucson, AZ 85721}

\begin{abstract}

We present evidence that the incidence of active galactic nuclei (AGNs)
and the distribution of their accretion rates do not depend on the
stellar masses of their host galaxies, contrary to previous studies.
We use hard ($2-10$ keV) X-ray data from three extragalactic fields
(XMM-LSS, COSMOS and ELAIS-S1) with redshifts from the Prism 
Multi-object Survey to identify \amend{242} AGNs with
$L_\mathrm{2-10\; keV}=10^{42-44}$ erg s$^{-1}$ within a parent sample 
of $\sim$25,000 galaxies at $0.2<z<1.0$ over
$\sim3.4$~deg$^{2}$ and to $i\sim23$.  We find that although the fraction of galaxies
hosting an AGN at fixed X-ray luminosity rises strongly with stellar
mass, the \emph{distribution} of X-ray luminosities is independent of
mass.  Furthermore, we show that the probability that a galaxy will
host an AGN can be defined by a universal Eddington ratio distribution
that is \emph{independent} of the host galaxy stellar mass and has a
power-law shape with slope $-$\amend{0.65}.  These results demonstrate that AGNs
are prevalent at all stellar masses in the range
$9.5<\log \mathcal{M}_*/\mathcal{M}_\odot<12$ and that the same
physical processes regulate AGN activity in all galaxies in this stellar mass range.
While a higher AGN fraction may be observed 
in massive galaxies, this is a selection effect related to the
underlying Eddington ratio distribution.  We also find that the AGN
fraction drops rapidly between $z\sim1$ and the present day and is
moderately enhanced (factor$\sim2$) in galaxies with blue or green
optical colors.  Consequently, while AGN activity and star formation
appear to be globally correlated, we do not find evidence that the presence of an AGN
is related to the quenching of star formation or the color
transformation of galaxies.
\end{abstract}

\maketitle


\section{Introduction}
\label{sec:intro}

Supermassive black holes (SMBHs) are now thought to reside at the centers of most, if not all,
galaxies with central stellar bulges \citep{Kormendy95,Kormendy01}. 
These galaxies must have undergone periods of intense accretion
activity to build their black holes, during which they would be
observed as active galactic nuclei (AGNs).  
The mass of the SMBH is
known to be tightly correlated with both the luminosity
\citep{Magorrian98} and velocity dispersion
\citep{Ferrarese00,Gebhardt00,Tremaine02} of the stellar bulge,
indicating that the growth of the SMBH and the  
processes that build up the stellar mass of the host galaxy must be related, 
\amend{although it is unclear whether a causal connection between these processes is required \citep[see][]{Peng07,Jahnke11}.}
In addition, the star formation rate density as a function of redshift is found to 
correlate with the black hole accretion rate density out to high redshifts \citep[e.g.,][]{Silverman08,Aird10}, 
indicating that these processes may be 
\amend{controlled}
by a common mechanism.
Indeed, a number of studies have proposed that AGNs provide feedback
that quenches star formation in their host galaxies and is crucial to control the
transformation of blue, star-forming galaxies to red, passively evolving systems
 \citep[e.g.,][]{Croton06,Hopkins07,Schawinski07}, although 
 there remains a lack of direct observational evidence of this process.
  Nevertheless, what triggers and regulates AGN activity
 and how this relates to the co-evolution of AGNs and their host galaxies remain key, open
 questions in extragalactic astrophysics.

To understand how AGNs and galaxies co-evolve we must study the types of galaxies that tend to host AGNs. 
Galaxies can be separated into a variety of classifications, many of which have significant overlap.
Many studies have shown that the positions of galaxies in the optical color--magnitude diagram are strongly bimodal, falling into two distinct regions associated with passively evolving galaxies on the ``red sequence" and star-forming galaxies in the ``blue cloud" \citep[e.g.,][]{Blanton03,Baldry04,Bell04}; a narrow ``green valley" separates these two populations.
The total stellar mass of a galaxy, \Mstel, is also a key parameter that tracks the net result of star formation activity throughout a galaxy's lifetime, 
as well as the hierarchical build up of galaxies through mergers.
While both the red and blue galaxy populations span a wide range of stellar masses, red galaxies tend to dominate the population at higher stellar masses, whereas blue galaxies generally have lower stellar masses.
Over cosmic time the total combined mass of galaxies on the red sequence is found to increase, indicating that star formation must be quenched in blue galaxies, 
\amend{which subsequently allows their stellar populations to evolve passively and transfers the galaxies to the red sequence \citep{Bell04,Faber07,Brown07}.}
Morphology is also closely correlated with galaxy color and stellar mass---massive, red galaxies are generally early-type, bulge-dominated systems, whereas blue galaxies tend to have late-type morphologies with substantial disk components \citep[e.g.,][]{Blanton09}. 

Which of these galaxies host AGNs? 
\amend{Given the correlation between the overall AGN accretion and star formation rate densities \citep{Silverman08,Aird10},}
we might expect to find AGNs predominantly \moreamend{within blue star-forming galaxies, which tend to have low stellar masses.} 
A number of studies show that this is not the case---the fraction of galaxies hosting AGNs increases in higher stellar mass, bulge-dominated galaxies---although the subject remains under much debate.
\citet{Kauffmann03} found that narrow-line AGNs (identified by their emission-line ratios) are preferentially in massive, bulge-dominated galaxies (\Mstel$\gtrsim10^{10}$\Msun) in the local universe. 
\amend{\citet{Kauffmann03} also found evidence for recent star formation in the host galaxies of the more luminous AGNs, although the host galaxies of the lower luminosity AGNs generally had older stellar populations, similar to most early-type galaxies \citep[see also][]{Kauffmann07}.
}
The fraction of galaxies that host radio-loud AGNs is even more strongly dependent on stellar mass,  rising rapidly from $<0.01$\% at \Mstel$\approx 10^{10}$ \Msun\ to $>30$\% at \Mstel$\gtrsim 5\times 10^{11}$\Msun \citep[]{Best05};
it has long been established that radio-loud AGNs are primarily found in massive, elliptical galaxies \citep[e.g.,][]{Matthews64,Dunlop03}.
\citet{Schawinski07} studied the optical colors of elliptical galaxies in the local universe and found that those hosting narrow-line AGNs were generally on the red sequence\amend{; these}
results suggest that star formation and AGN activity do not occur at the same time, which could be explained by the onset of AGN activity  (possibly during an early, highly obscured phase of the AGN lifetime) being responsible for quenching star formation activity within galaxies and transferring them from the blue cloud to the red sequence \citep[e.g.,][]{Hopkins06b}. 
\amend{
However, direct observational evidence of the AGN feedback process is lacking.
A detailed study of the prevalence of AGNs as a function of stellar mass and galaxy color is required to shed light on the relationship between the growth of galaxies, the progress of star formation, and the accretion activity of the central SMBH. 
}

Deep X-ray surveys have enabled efficient identification of obscured and low-luminosity AGNs where the host galaxy dominates the optical light, allowing studies of the host galaxy properties to be pushed out to much higher redshifts \citep[$z\lesssim4$;][]{Xue10}.
Various studies have shown that the fraction of galaxies hosting X-ray AGNs 
rises strongly with increasing \Mstel\ at all redshifts from $z\sim0.4-4$ \citep[e.g.,][]{Alonso-Herrero08,Brusa09b,Xue10}, similar to the optical narrow-line AGNs identified in the local universe \citep{Kauffmann03}. 
\citet{Nandra07b} investigated the positions of X-ray AGN hosts in the optical color--magnitude diagram at $z\sim1$ and found that X-ray-selected AGNs are predominantly in luminous galaxies 
but avoid the strongest star-forming galaxies and thus lie in the reddest part of the blue cloud, in the green valley, or on the red sequence. 
These results indicate that AGN activity may be associated with the transformation of blue galaxies and the quenching of star formation, indicating that AGN feedback could potentially play a key role in this process.
Similar results were found by \citet{Coil09} for a larger sample of X-ray AGNs, and the same pattern appears to hold at lower redshifts \citep{Hickox09,Georgakakis11}. 
However, \citet{Silverman09} and \citet{Xue10} have demonstrated the importance of stellar-mass selection effects, which may bias prior work.
\citet{Silverman09} found that X-ray AGN activity was associated with ongoing star formation, and thus the AGN fraction is higher in the blue cloud. 
\citet{Xue10} found that the AGN fraction rises strongly with increasing \Mstel, as described above, but when considering samples matched in stellar mass there was no evidence for specific clustering of AGN host galaxies on the color--magnitude diagram. 
\citet{Cardamone10b} also studied the relation of AGN host galaxy masses and colors; they found that the colors of AGN host galaxies of a given stellar mass followed the same bimodality as the non-AGN galaxy population, after correcting for the effects of dust. 
Thus, despite much recent progress, it remains unclear as to whether there is an association between AGN activity and the color transformation of galaxies, and how this relates to the predominance of AGNs in massive galaxies.

Further insight into the processes that trigger AGNs and how these relate to the evolution of galaxies may be possible by directly probing the distribution of AGN activity, the active growth of SMBHs via accretion that produces a luminous AGN.
A number of studies have  constrained the shape and evolution of the AGN luminosity function, which traces AGN activity, over a wide redshift range
\citep[e.g.,][]{Ueda03,Ebrero09,Yencho09,Aird10,Assef11}.
The luminosity function has a steep double power-law shape with a break at a characteristic luminosity, $L_*$.
The rapid decline in the AGN accretion rate density since $z\sim1$ is a result of a downward shift in $L_*$ for the overall population, but the luminosity function retains \amend{essentially} the same shape \citep{Barger05,Aird10,Assef11}.
This indicates that the processes responsible for triggering AGNs, fueling black hole growth and controlling the lifetimes and duty cycles of AGNs are essentially the same but are moving towards generally less luminous systems between $z\sim1$ and the present day. 
However, few observational studies have attempted to simultaneously study the distribution of AGN activity as traced by the AGN luminosity and the prevalence of AGNs as a function of the host galaxy properties such as color and stellar mass.

Understanding both the predominance of AGNs in massive galaxies and any correlation with host galaxy colors is vital to reveal the role and effect of AGNs on the evolution of galaxies. 
\amend{In addition, to constrain the processes that trigger AGNs and drive black hole growth we must track the evolution of AGN accretion rates and the masses of their central SMBHs, although direct measurement of these quantities is extremely difficult, if not impossible for large samples of distant AGNs and galaxies.}
\amend{In this paper, we use the X-ray luminosity as a tracer of the AGN accretion rate and investigate the relationship with the total stellar mass and colors of galaxies. 
While the relationship between these observed properties and the underlying physical parameters remains uncertain, careful investigation including consideration of observational biases and selection effects is vital to reveal and constrain the connection between the growth of galaxies and their SMBHs.
Our study focuses on } intermediate redshifts ($z<1$), a key time period of the universe over which star formation and AGN activity have rapidly declined from the peak at $z\sim1-2$.
We use the PRIsm MUlti-object Survey 
\citep[PRIMUS;][Cool et al., in preparation]{Coil10}, a recently completed faint-galaxy survey that used a prism to obtain low-resolution spectra and determine 
redshifts for $\sim 120,000$ galaxies over $\sim 9 $ \degsq\ with $z\sim 0.2-1.2$, providing the largest currently available spectroscopic sample of galaxies at these redshifts.
X-ray data are available for $\gtrsim 3$ \degsq\ of PRIMUS
from a variety of \textit{Chandra} and \textit{XMM-Newton} survey programs. 
We use these X-ray data to identify AGNs within the PRIMUS sample, enabling us to efficiently select AGNs down to low luminosities and with moderate obscuration.
For such sources the optical light probes the host galaxies, and we can compare properties with the extremely large non-AGN 
galaxy sample provided by PRIMUS.
Section \ref{sec:data} briefly describes our data, providing details on the compilation of X-ray source lists and matching to PRIMUS 
sources. 
Section \ref{sec:stellarmass} outlines our derivation of stellar masses and Section \ref{sec:sample} describes in detail the selection of our AGN and galaxy samples.
We investigate how the probability of a galaxy hosting an AGN is related to the stellar mass of the galaxy and the AGN X-ray luminosity in Section \ref{sec:frac} for two wide redshift bins. 
In Section \ref{sec:zevol} we determine the redshift dependence of the AGN fraction and perform a more 
robust analysis of the stellar mass and luminosity dependence that accounts for the evolution with redshift via a maximum-likelihood fitting approach.
In Section \ref{sec:eddratio} we show that the probability that a galaxy hosts an AGN is primarily determined by a power-law distribution of Eddington ratios with a slope of $-0.65$ and a normalization that depends on redshift but does \emph{not} depend on the stellar mass of the host galaxies.
We investigate the dependence of the AGN fraction on galaxy color in Section \ref{sec:col} and find a weak enhancement of AGNs in galaxies with bluer colors.
We discuss the implications of our results in Section \ref{sec:discuss}.
Section \ref{sec:summary} summarizes our results and overall conclusions.

Throughout this paper we adopt a flat cosmology with $\Omega_\Lambda=0.7$ and $h=0.7$. All magnitudes are on the AB system.


\section{Data}
\label{sec:data}

\subsection{PRIMUS}

PRIMUS is the largest faint-galaxy, intermediate-redshift survey performed to date and covers $\sim 9$ deg$^2$ in a total of seven different 
well-studied multiwavelength fields. We used the IMACS instrument on the Magellan I Baade 6.5m telescope with a slitmask and 
low-dispersion prism to obtain low-resolution ($\lambda/\Delta\lambda\sim40$) spectra for $\sim 2000$ objects \amend{with a single slitmask.
We generally observed two slitmasks for each pointing, allowing us to assign slits to a statistically complete sample down to $i\approx23$. 
At our faintest magnitudes ($i\sim22.5-23$) we sparse-sample galaxies with a weight of 0.3 if an object has potential slit collisions; this sparse-sampling ensures that PRIMUS is not dominated in number by the faintest galaxies.
We also applied a density-dependent sampling scheme to all objects with more than two nearby neighbors on the plane of the sky to reduce the effects of slit collisions in a well-controlled manner.
Overall we were able to assign slits to $\sim80$\% of galaxies at $i\lesssim22.5$.
Unresolved point sources were only retained if their optical and/or IR colors (or an X-ray detection for the Subaru/\textit{XMM-Newton} Deep Survye, SXDS, area of the XMM-LSS field) indicated they may contain an AGN; however, no additional prioritization was given to such sources in the targeting or slitmask design.
For full details of the survey design, targeting and a data summary see \citet{Coil10}.}

\amend{
To obtain redshifts we fit the low-resolution spectra and photometry with galaxy as well as BLAGN and stellar spectral 
templates. 
We supplement our low-resolution spectra with optical ground-based photometry, which is used in the redshift fitting as well as to derive $K$-corrections and stellar mass estimates (see Section 
\ref{sec:stellarmass} below). 
We classify the objects as stars, BLAGNs or galaxies based on the $\chi^2$ of the best fits.
Our redshift success rate is $\sim90$\% at bright magnitudes ($i<21$) and drops to $\sim70$\% at $i\sim22.5$, but has little dependence on galaxy color.
Our final redshift catalog contains $\sim 120,000$ robust redshifts with a precision of 
\amend{$\sigma_z/(1+z)=$0.005. (normalized median absolute deviation)}
For details of the data reduction, redshift fitting algorithm, redshift confidence and 
precision, and survey completeness see R. J. Cool et al. (in preparation).}

\subsection{X-ray data}
\label{sec:xray} 

X-ray data of varying depths from \textit{Chandra} and/or \textit{XMM-Newton} are available for five of our PRIMUS fields. 
For this paper, we have compiled publicly available X-ray point source catalogs in three of these fields: XMM-LSS, COSMOS and ELAIS-S1. 
\amend{Further details of the catalogs in the individual fields are given in Sections \ref{sec:xmm}, \ref{sec:cosmos} and \ref{sec:es1} below.
We note that these X-ray catalogs were compiled \emph{after} the PRIMUS observations had been completed. 
PRIMUS is a purely flux limited survey and the X-ray information had no bearing of the targeting of sources.}

In all our fields we have used the likelihood ratio matching technique \citep[e.g.,][]{Sutherland92,Ciliegi03,Brusa07,Laird09} to robustly identify the counterparts of each X-ray source in the PRIMUS targeting optical photometry \citep[usually the $i$ or $R$ band, see Table 3 in][]{Coil10}.
The likelihood ratio technique allows us to assign secure optical counterparts to the X-ray sources, accounting for the optical and X-ray positional uncertainties, the probability of a counterpart with a given magnitude, and the probability of a spurious match. Candidate counterparts are identified within 5\arcsec\ of an X-ray source position. 
While a 5\arcsec\ search radius is significantly larger than the typical X-ray positional uncertainties for on-axis \chandra\ data, it is appropriate for \xmm\ or off-axis \chandra\ data. 
As the likelihood ratio takes account of the positional uncertainties for individual sources, we can adopt a single search radius for all our fields and data.
Where multiple counterparts exist we choose the match with the highest likelihood ratio; we further restrict our matches to ``secure" counterparts with likelihood ratios $>0.5$.

We match these optical counterparts to sources actually targeted by PRIMUS
\amend{and inspected the PRIMUS spectra of every X-ray source by eye.
These manual checks allow us to confirm the PRIMUS classifications and redshifts, identify spectra with clear features (such as broad lines) that may not be well-fitted by the automated procedure, select between multiple possible redshift solutions and identify catastrophic data issues. 
Based on these manual checks we altered the redshifts and classifications of around 20\% of our X-ray sources, mainly reclassifying sources as BLAGNs.}

\amend{
After completing our manual checks, we compiled prior estimates of the redshifts of our sources based on high-resolution spectroscopy from other surveys. 
In the COSMOS field we use the zCOSMOS catalog \citep{Lilly09} as well as redshifts from various spectroscopic campaigns following up XMM-COSMOS sources, as presented in the recent compilation by \citet{Brusa10}.
In the XMM-LSS we use the compilation of spectroscopic redshifts in the UDS area\footnote{http://www.nottingham.ac.uk/astronomy/UDS/data/data.html}, which includes redshifts from \citet{Smail08}, \citet[C. ][in preparation]{Simpson11} and \citet[M. ][in preparation]{Akiyama11}.
In the ELAIS-S1 field we use the spectroscopic redshifts of X-ray sources presented by \citet{Feruglio08}.
Overall this provides high resolution spectroscopic redshifts for 270 of the X-ray sources within PRIMUS.
Approximately 50\% of these sources are classified as BLAGNs.
For the sources with galaxy classifications we determine the precision of the robust PRIMUS redshifts to be $\sigma_z/(1+z)=0.008$. 
The outlier rate depends on magnitude: 
at $i<22$, corresponding to 85\% of our final sample (see Section \ref{sec:sample} below), the outlier rate (fraction with $\delta z/(1+z)>0.1$) is 
4.8\%; at fainter magnitudes the outlier rate may rise to $\sim30$\%, but is fairly poorly determined due to the lack of sources reliable high-resolution spectroscopic follow-up to this depth. 
We note that the outlier rate for the X-ray detected galaxy population is slightly higher than determined for the overall PRIMUS galaxy sample \citep[2.0\%: see][]{Coil10} but is low enough that we do not expect a significant impact on our results.
}

\subsubsection{XMM-LSS}
\label{sec:xmm}

In the XMM-LSS field we use the catalog of \citet{Ueda08} from the SXDS, based on seven deep \xmm\ pointings (1.14 deg$^2$) that overlap our PRIMUS observations.
We supplement this with the point source catalog published in \citet{Pierre07}, which utilizes the first 45 pointings (5.5 deg$^2$) of the XMM-LSS X-ray survey, although the overlap with PRIMUS is limited.
Our combined catalog contains 4466 unique sources down to hard band (2--10 keV) fluxes of $f_{2-10 \mathrm{keV}}\sim2\times10^{-15}$ \ergscm, 1073 of which fall within the area covered by PRIMUS masks, and 617 of which have secure optical counterparts. 
We obtained robust redshifts and classifications for \amend{266} of these X-ray sources with PRIMUS over a total area of $1.78$ deg$^2$ that is  covered by both our PRIMUS masks and the X-ray data. 
We note that an additional $\sim 1$ deg$^2$ of our PRIMUS area is covered by X-ray data from the XMM-LSS survey that was not included in the \citet{Pierre07} catalog, and is thus not used here.

\subsubsection{COSMOS}
\label{sec:cosmos}

The entire 2 deg$^2$ COSMOS field has been observed with \xmm\ \citep{Hasinger07} to depths of $f_\mathrm{2-10 keV}\sim3\times10^{-15}$ \ergscm; much deeper \chandra\ data reaching $f_\mathrm{2-10 keV}\sim8 \times10^{-16}$ \ergscm\ were also obtained for the central $\sim 0.9$ deg$^2$ \citep{Elvis09}.
We merged the \xmm\ \citep{Cappelluti09} and \chandra\ \citep{Elvis09} point source catalogs, and removed duplicates by first matching bright sources ($f_\mathrm{0.5-10 keV}>10^{-14}$ \ergscm) to correct for any overall offset between the two catalogs, and then searching for the counterparts of the \xmm\ sources within the \chandra\ data using a 5\arcsec\ search radius. For matched sources we adopt the positions and fluxes from the \chandra\ catalog. We find that there are $\sim 70$ \xmm\ sources that fall within the approximate area observed with \chandra\ but lack \chandra\ counterparts, consistent with \citet{Elvis09} and attributed to areas or small gaps with low \chandra\ exposure, or some expected fraction of spurious sources. We retain all unmatched \xmm\ sources in our final merged catalogs; spurious sources are very unlikely to have secure optical counterparts above our PRIMUS magnitude limits so will not significantly contaminate our sample. Our merged catalog contains 1232 unique hard X-ray sources over the 1.07 \degsq\ with both PRIMUS and X-ray coverage; 307 of these have robust optical counterparts, were targeted by PRIMUS, and have robust redshift measurements and classifications from the PRIMUS data. 

\subsubsection{ELAIS-S1}
\label{sec:es1}

\citet{Puccetti06} presented a point source catalog from a mosaic of four partially overlapping \xmm\ pointings in the ELAIS-S1 field that reach depths of $f_\mathrm{2-10 keV}\sim2\times10^{-15}$ \ergscm. This provides X-ray data coverage for 0.52 \degsq\ of our total 0.9 \degsq\ PRIMUS area, although PRIMUS has little overlap with the deepest portions of the ELAIS-S1 X-ray data.
One hundred and twenty-nine hard X-ray sources fall within the PRIMUS area of which 105 have secure optical counterparts. Robust redshifts and classifications for 59 sources are provided by PRIMUS.

\begin{table*}
\caption{Details and Source Numbers for our PRIMUS/X-ray fields
\label{tab:primusxtab}
}
\begin{center}
\begin{tabular}{l c c c c c c c c c c c c}
\small
\vspace{-5mm}\cr
\hline
\hline
\vspace{-2 mm} \cr
Field Name & Area (deg$^2$)& \multicolumn{7}{c}{X-ray Source Numbers} & Number of Galaxies \\
 & (1) & (2) & (3) & (4) & (5) & (6) & (7) & (8) & (9) \\
\vspace{-2 mm} \cr
\hline 
\vspace{-2 mm} \cr
    XMM-LSS &    1.78 & 1073 & 617 & 410 & \amend{266} & \amend{116} & \amend{102} &  \amend{83}     & \amend{12523} \\
    COSMOS  &   1.07  & 1232 & 914 & 514 & \amend{307} & \amend{161} & \amend{144} & \amend{138}   & \amend{10114} \\
     ELAIS-S1 &    0.52 &   129 & 105  &  82  &  \amend{59}  &   \amend{25}  &  \amend{22}  &   \amend{21}    & \amend{3079} \\
\vspace{-2 mm} \cr
\hline                      
\vspace{-2 mm} \cr   
    Total          & 3.36 & 2434 & 1636 & 1006 & \amend{632} & \amend{302} & \amend{268} & \amend{242}  &   \amend{25716} \\
\vspace{-2 mm} \cr    
\hline
\vspace{-2 mm} \cr    
\multicolumn{10}{p{12cm}}{Column 1:  area with joint X-ray and PRIMUS coverage.}\\
\multicolumn{10}{p{12cm}}{Column 2: total number of hard-band detected X-ray sources in X-ray/PRIMUS area.}\\
\multicolumn{10}{p{12cm}}{Column 3: number with robust optical counterparts.}\\
\multicolumn{10}{p{12cm}}{Column 4: number of counterparts targeted by PRIMUS.}\\
\multicolumn{10}{p{12cm}}{Column 5: number with robust redshifts ($Q\ge 3$) from PRIMUS.}\\
\multicolumn{10}{p{12cm}}{Column 6: number with robust redshifts in the range $0.2<z<1.0$ and X-ray luminosities in the range $42<\log \lx<44$.}\\
\multicolumn{10}{p{12cm}}{Column 7: number with robust redshifts, $0.2<z<1.0$, $42<\log \lx<44$ and galaxy classifications.}\\
\multicolumn{10}{p{12cm}}{Column 8: final number with robust redshifts, $0.2<z<1.0$, $42<\log \lx<44$, galaxy classifications and \Mstel above the redshift-dependent stellar mass limit determined in Section \ref{sec:stellarmasslim}.}\\
\multicolumn{10}{p{12cm}}{Column 9: total number of galaxies in sample with robust redshifts, $0.2<z<1.0$ and \Mstel\ above the redshift-dependent stellar mass limit.}\\
\end{tabular}
\end{center}
\end{table*}


\section{Stellar mass estimates}
\label{sec:stellarmass}

We estimate stellar masses by fitting the observed
optical/near-infrared spectral energy distributions (SEDs) of the
galaxies in our sample.
\amend{In the XMM-LSS field we construct observed SEDs based on the $u^*g'r'i'z'$ imaging from Megacam on the Canada-France Hawaii Telescope \citep[CFHT: ][]{Boulade03} obtained as part of the CFHTLS-Wide survey\footnote{http://www.cfht.hawaii.edu/Science/CFHLS};
we use the photometric catalogs released by the CFHTLS Archive Research Survey \citep[][]{Erben09}.
In the \moreamend{COSMOS} field we use
$B_J V_J g^+ r^+ i^+ z^+$ imaging obtained with the Suprime-Cam instrument on the Subaru 8.2m telescope \citep{Miyazaki02}, $u^*$ and $i^*$ imaging from Megacam on the CFHT, and $K_s$ imaging from the Wide-field Infrared Camera at the CFHT \citep{McCracken10}; 
photometry in all these bands is obtained from the public multiwavelength photometric catalog released in 2009 April\footnote{http://irsa.ipac.caltech.edu/data/COSMOS}. 
In the ELAIS-S1 field we use $BVR$ imaging obtained with the Wide Field Imager at the 2.2 m La Silla
ESO-MPI telescope \citep{Berta06} and $Iz$ imaging based on observations with the Visible Multi Object
Spectrograph camera at the Very Large Telescope \citep{Berta08}.
}

\amend{We perform the SED fitting} using {\tt iSEDfit} (see J. Moustakas et al., in
preparation, for complete details).  Briefly, {\tt iSEDfit} is a Bayesian
SED-fitting code that compares the observed photometry against a large
Monte Carlo suite of model SEDs spanning a wide range of age,
metallicity, dust content, and star formation history \citep[see,
  e.g., ][]{Kauffmann03b, Salim07}.  By marginalizing over all other
parameters, we construct the posterior probability
distribution of stellar masses, $p(\mstel)$, for each galaxy, 
encapsulating both the
photometric uncertainties and the physical degeneracies among
different models.  We take the median of the $p(\mstel)$ distribution as
our best estimate of the stellar mass.

We construct our models using the \citet{Bruzual03} stellar population
synthesis models, assuming the \citet{Chabrier03} initial mass
function (IMF) from 0.1 to 100~\Msun.  We assume a uniform prior on
stellar metallicity in the range $0.004<Z<0.05$, and smooth,
exponentially declining star formation histories, $\psi(t)\propto
\exp^{-\gamma t}$, with $\gamma$ drawn uniformly from the interval
$[0.01,1]$~Gyr$^{-1}$.  The time since the onset of star formation is
given by $t$, which we draw from a uniform distribution ranging from
$0.1$ to $13$~Gyr.  However, we apply an additional prior that disallows
ages older than the age of the universe at the redshift of each
galaxy.  Finally, we adopt the time-dependent dust attenuation law of
\citet{Charlot00} in which stars older than $10$~Myr are attenuated
by a factor $\mu$ less than younger stars.  \amend{We draw the
  $V$-band optical depths from an order two Gamma distribution that
  peaks around $A_{V}\approx1.2$~mag, with a tail to
  $A_{V}\approx6$~mag, and $\mu$ from an order four Gamma distribution
  that ranges from zero to unity centered on a typical value
  $\langle\mu\rangle=0.3$ \citep[see][]{Charlot00, Wild11}.}

Adopting different prior distributions---for example, superposing
stochastic bursts on otherwise smooth star formation histories, or
adopting different metallicity or reddening priors---changes our
stellar masses by $\lesssim0.1$~dex in the mean, and by typically less
than a factor of two for individual objects (see J. Moustakas et al., in
preparation, for details).  Given that our sample spans more than two orders
of magnitude in stellar mass, we do not expect these uncertainties to
affect our conclusions.  The largest potential systematic effect is
due to our assumed IMF; however, that will simply shift all our
stellar mass estimates by an equal amount.  


\section{Sample selection}
\label{sec:sample}

In this paper, we wish to investigate the fraction of X-ray AGNs within the PRIMUS galaxy sample as a function of both the galaxy properties (\Mstel, color) and the X-ray luminosity. Our galaxy sample is restricted to
(1) sources within the 3.37 \degsq\ of PRIMUS with X-ray coverage;
(2) robust redshift measurements ($Q\ge 3$,  see \citealt{Coil10} and R. J. Cool et al. in preparation for details);
and (3) galaxy classifications -- objects where the PRIMUS spectrum is best fit by a BLAGN template are excluded (see Section \ref{sec:blagn} below).
Our X-ray data are then used to identify AGNs \emph{within this galaxy sample}. 

\amend{Our sample includes both primary PRIMUS targets, with well defined targeting (sparse and density-dependent) weights, and a small number of secondary targets that are drawn from the same parent sample of sources but do not have a well-defined weight.
We experimented with further restricting our sample to only primary targets, and applying corrections based on the targeting weights. 
This reduces our sample of X-ray detected AGNs but did not otherwise significantly alter the results presented in the rest of this paper. 
This indicates that the targeting incompleteness of the PRIMUS sample affects our galaxy sample and X-ray detected AGN sub-sample roughly equally;  the small scale collisions that determine the targeting weights do not impact our derived fraction of AGNs. 
We therefore retain both primary and secondary targets to maximize our sample sizes.}

\amend{When a reliable high-resolution spectroscopic redshift is available from one of the catalogs described in Section \ref{sec:xray} above, we adopt this value rather than the PRIMUS redshift.
In the COSMOS field we adopt reliable redshifts from the zCOSMOS-10k bright sample \citep{Lilly09} with a confidence flag of 3 or 4; we do not use the additional redshifts of X-ray sources from \citet{Brusa10} in place of our PRIMUS redshifts due to the heterogeneity of this catalog and the lack of the redshift confidence flags for individual sources.}
This provides updated redshifts for \amend{1643} sources in our three PRIMUS fields with X-ray coverage.

Table \ref{tab:primusxtab} gives the numbers of galaxies and X-ray sources in our three fields that satisfy our various selection cuts.

\subsection{X-ray selection}
\label{sec:xrayselection}

We restrict our AGN identifications to hard-band (2-10 keV) detections. 
Hard-band selection identifies a sample that includes both X-ray-unabsorbed and moderately absorbed AGNs, although it will miss the most heavily absorbed, Compton-thick sources. Soft (0.5--2 keV) or full (0.5--10 keV) band selection, by comparison, will be biased toward the detection of unabsorbed sources (equivalent hydrogen column densities $N_\mathrm{H}\lesssim10^{22}$\cmsq). 
Hard-band selection also ensures that we can accurately measure a rest-frame 2--10 keV X-ray luminosity based on the observed hard-band flux; we calculate luminosities assuming a photon index $\Gamma=1.9$. 
A disadvantage of hard band selection is the reduction in the size of our X-ray detected sample due to the lower sensitivity of X-ray observatories at higher energies.
We further limit our AGN sample to $L_\mathrm{X}>10^{42}$ \ergs\ to ensure that our measured X-ray luminosity can be associated with an AGN (rather than star formation), and
$L_\mathrm{X}<10^{44}$ \ergs\ to exclude the brightest AGNs that may contaminate the host galaxy light and to reduce the effects of our selection against unobscured AGNs (see Section \ref{sec:blagn} below). 
Very few of the X-ray detected AGNs \amend{within our galaxy sample} lie above $\lx=10^{44}$ \ergs\ so this cut has little effect on our results.
Hard band selection also ensures our AGN selection criteria are not indirectly affected by the host galaxy; conversely, if we choose to include soft-band-selected sources we could be biased by absorption effects within the host galaxy

\begin{figure}
\begin{center}
\includegraphics[width=0.45\textwidth]{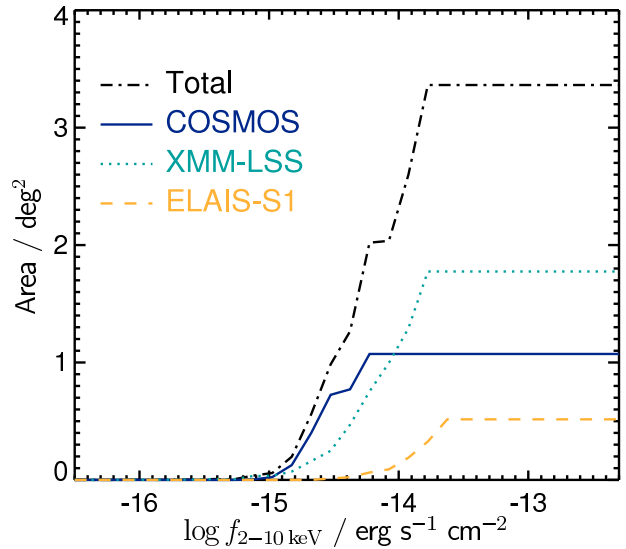}
\end{center}
\caption{
X-ray area curves for each of our fields and total combined sample derived by comparing the number of sources in our compiled catalogs at a given hard (2-10 keV) X-ray flux with predictions based on the $\log N-\log S$ relations of \citet{Georgakakis08}. These allow us to correct for incompleteness due to the varying depths of our X-ray data, both between our fields and within an individual field. 
}
\label{fig:acurve}
\end{figure}

The X-ray selection of AGNs will be subject to significant incompleteness in the data that we \emph{must} correct for. 
The depth of our X-ray data varies  between our fields and within a single field due to the differing X-ray observatories, the varying  vignetting and point spread functions within a pointing, and the different exposure times of the pointings. Our entire survey area is sensitive to bright fluxes, $f_\mathrm{2-10 keV}\gtrsim 2 \times 10^{-14}$ \ergscm, thus we will identify all AGNs with $L_\mathrm{X}\gtrsim10^{44}$ \ergs\ out to $z=1$ (with the exception of Compton-thick sources). 
However, this luminosity cut would severely reduce our sample size and we are mainly interested in studying the moderate luminosity ($L_\mathrm{X}\approx10^{42-44}$ \ergs) AGN  population that dominates the luminosity density at $z\lesssim1.0$ \citep{Aird10} and where we are able to probe the properties of the host galaxy (as for the majority of this population the AGNs are moderately obscured).
Instead of the strict flux cut, we apply completeness corrections to our derived AGN fractions based on the X-ray area (sensitivity) curves.
We calculate the area curves by comparing the number of hard X-ray detected sources in bins of 0.1 dex from our parent point source catalogs (regardless of optical counterparts or PRIMUS targeting) that fall within our PRIMUS window with the predicted numbers as a function of flux based on the X-ray $\log N -\log S$ relation of \citet{Georgakakis08}. 
We fix the area curve at the total area coverage above the lowest flux where $>99$\% of the predicted sources are detected.
This prevents small number statistics adding noise to our area curves at the highest fluxes.
Area curves derived in this way are shown in Figure \ref{fig:acurve}.

Two different approaches to incorporate this completeness information are described in Sections \ref{sec:frac} and \ref{sec:zevol} below. Figure \ref{fig:lx_vs_z} shows the distribution of X-ray luminosities for our X-ray detected sample, and the bins applied in Section \ref{sec:frac}.
The effects of the X-ray incompleteness are clearly seen in the lack of lower \Lx\ objects at high redshifts.

\begin{figure}
\begin{center}
\includegraphics[width=0.45\textwidth]{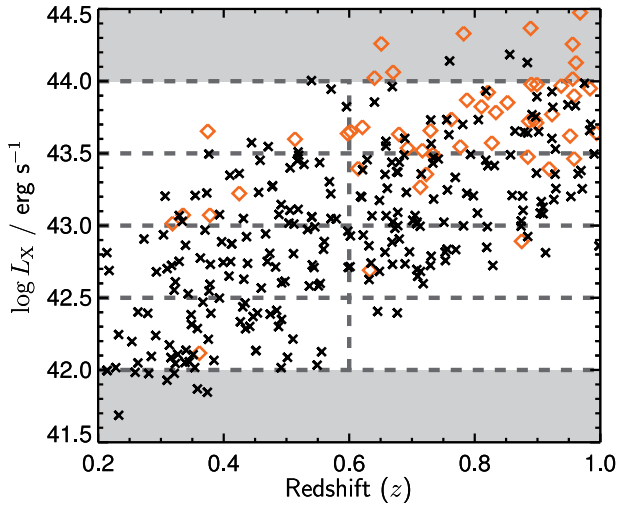}
\end{center}
\caption{
X-ray luminosity (2--10 keV) vs. redshift for the X-ray-detected sub-sample of galaxies (crosses) and X-ray detected broad-line AGsN (red diamonds). Dashed lines indicate the bins used in Section \ref{sec:frac}.
We restrict our sample to moderate-luminosity AGNs ($42<\log L_\mathrm{X}<44$) below the break in the X-ray luminosity function.
Our sample suffers from incompleteness due to the varying X-ray flux limits over our survey, which is particularly severe for $\log L_\mathrm{X}\lesssim 42.5$ at higher redshifts; this is fully described by our X-ray area curves (Figure \ref{fig:acurve}) and accounted for by our X-ray completeness corrections described in Sections \ref{sec:frac} and \ref{sec:zevol}.
We also exclude the broad-line AGNs, which are mainly found at higher X-ray luminosities ($\log \lx\gtrsim43.5$).
}
\label{fig:lx_vs_z}
\end{figure}

\begin{figure}
\begin{center}
\includegraphics[width=0.5\textwidth]{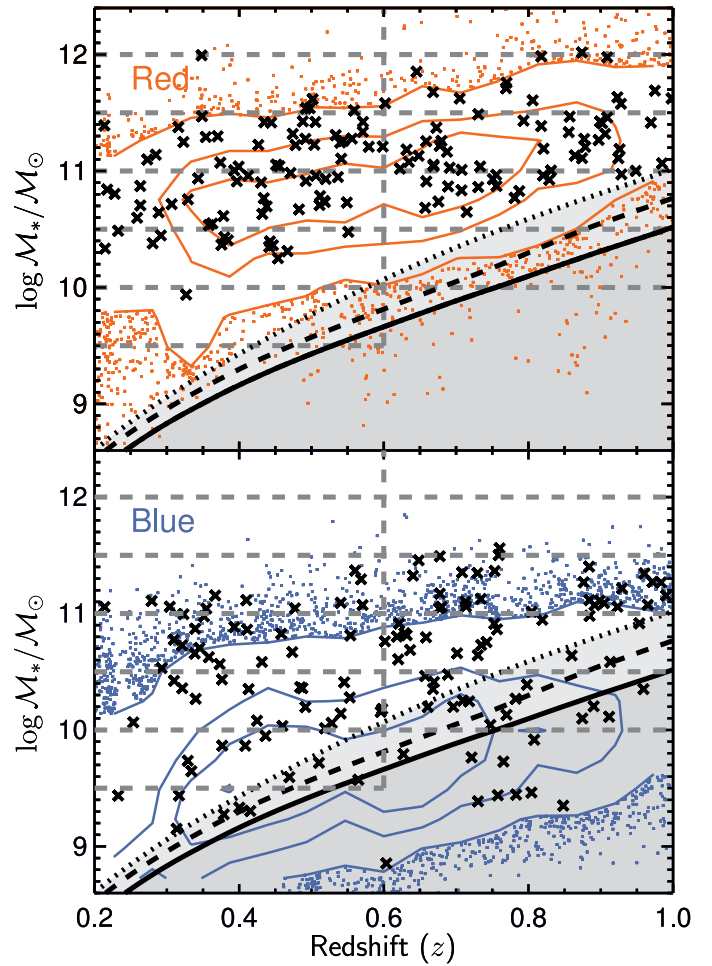}
\end{center}
\caption{
Stellar mass vs. redshift for our PRIMUS galaxy sample, divided into red (\textit{top}) and blue (\textit{bottom}) galaxies according to Equation \ref{eq:coldiv}. Contour levels enclose 30\%, 60\% and 90\% of the galaxy sample. X-ray-detected sources are shown by crosses. 
We define redshift-dependent stellar mass limits for the COSMOS (solid black curve), XMM-LSS (dashed black curve), and ELAIS-S1 (dotted black curve) fields based on the mass-to-light ratio of a maximally old stellar population (see Section \ref{sec:stellarmasslim}) and the limiting targeting magnitude in each field.
We cut both the red and blue galaxy samples using these mass limits; for blue galaxies we are able to probe to lower masses but we wish to limit any color-dependent incompleteness effects and thus apply this stricter cut.
The dashed gray lines indicate the bins used in Section \ref{sec:frac}
}
\label{fig:mass_vs_z}
\end{figure}

\subsection{Stellar mass limits}
\label{sec:stellarmasslim}

The galaxy sample will be subject to a number of incompletenesses due to the targeting and redshift success rates that will depend on the optical apparent magnitude and redshift, and thus indirectly depend on stellar mass.
However, the AGNs will be subject to the \emph{same} incompleteness effects (provided that the optical light is dominated by the host galaxy and not the AGN).
Thus the fraction of AGNs \emph{within} a sample of PRIMUS galaxies for a limited stellar mass and redshift range should reflect the true fraction of all galaxies that host AGNs at that stellar mass and redshift.
Nevertheless, we must set a lower limit on the stellar masses of galaxies in our sample to limit strong incompleteness effects that could depend on the host galaxy color and could bias our measurements of the AGN fraction.
First, we classify the galaxies as blue or red if they lie above or below the following relation that defines the ``green valley", 
\begin{equation}
u-g =  0.671 -0.031 M_g - 0.065 z
\label{eq:coldiv}
\end{equation}
where $u-g$ and $M_g$ are the rest-frame colors and absolute magnitudes \citep[derived using \kcorrect;][]{Blanton07} and $z$ is the redshift from PRIMUS. 
This cut is determined by identifying the peaks in $u-g$ color that correspond to the red sequence in 0.5 mag wide bins (in the range $-21<M_g<-19$) for galaxies with PRIMUS redshifts $0.4<z<0.7$.
A cut is made close to the minimum of the distributions and Gaussian fits are used to accurately determine the positions of these peaks.
A linear fit is performed to determine the slope of the red sequence as a function of $M_g$. 
We then take all galaxies in bins of $\Delta z=0.1$ over the range $0.4<z<0.7$ 
and adjust this slope blueward until it passes through the minimum corresponding to the green valley. 
We fit this color offset as a function of redshift, which allows us to account for the overall color evolution of the galaxy population.
Finally we identify the overall offset of the green valley for all galaxies at $0.2<z<1.0$ after accounting for the redshift evolution and magnitude dependence.

In Figure \ref{fig:mass_vs_z}, we show the distribution of stellar masses as a function of redshift for our red and blue galaxy samples and the X-ray detected sub-samples. 
We define a redshift-dependent stellar mass limit for each field above which strong, color-dependent incompleteness effects should be minimized. 
To define this limit we take a maximally old (i.e., $\tau=0$) simple stellar population with a formation redshift of $z=5$ and determine mass-to-light ratios (for the targeting optical bandpass) as a function of redshift. 
We use these mass-to-light ratios to convert our targeting magnitude limits in each field to an effective stellar mass limit as a function of redshift.
These stellar mass limits are shown by the black curves in Figure \ref{fig:mass_vs_z}. 
Above this mass limit our sample should be representative of the entire galaxy population as no stellar population should be redder than our maximally old template. 
We apply this cut to both the red and blue galaxy samples. 
We are able to detect blue galaxies and indeed some red galaxies down to much lower masses, but above this mass cut we will detect \emph{both} red and blue galaxies. 
This ensures that we can accurately measure the probability of a galaxy of a given stellar mass hosting an AGN, and will not be biased if AGNs preferentially reside in red or blue host galaxies. In section \ref{sec:cmd} we will investigate the positions of AGN host galaxies in the color-magnitude diagram in greater detail.

The dashed grey lines in Figure \ref{fig:mass_vs_z} show the redshift and mass bins used to derive the AGN fraction in section \ref{sec:frac}. 
The mass cut passes through our lowest stellar mass bins, reducing the number of galaxies in that bin. 
The number of detected AGNs should also be reduced by the same amount, and thus we will accurately measure the AGN fraction, assuming the AGN fraction does not strongly depend on mass or redshift \emph{within that bin}. In Section \ref{sec:zevol} we use a maximum likelihood approach to fit functional forms to the AGN fraction. The same mass cut is applied, but no binning is required and thus any biases associated with the assumption of a constant fraction within a bin are avoided.

\subsection{Broad-line AGNs}
\label{sec:blagn}

Our final selection criterion requires a galaxy classification based on the PRIMUS template fit.
This will exclude unobscured AGNs that exhibit broad emission lines in the optical spectrum.
We classify objects based on the relative $\chi^2$ values for the best-fit BLAGN and galaxy templates (see R. J. Cool et al., in preparation, for details).
This ensures we only exclude objects that are significantly better fit by a BLAGN template.

The BLAGN cut is essential as we can only accurately measure stellar masses for AGNs where the host galaxy dominates the optical light. 
This does introduce an additional incompleteness that we do not attempt to correct, and thus our study of the AGN fraction is only applicable to the ``obscured" population. 
Our definition of ``obscured'' means simply lacking strong broad lines in the optical spectrum; this could be due to intrinsic obscuration close to the SMBH (the torus), obscuration within the host galaxy \emph{or} dilution of the light from a weak AGN within a substantially brighter host galaxy. 
Figure \ref{fig:lx_vs_z} shows that the BLAGNs are mainly found at high X-ray luminosities, $\lx>10^{43.5}$ \ergs, consistent with the increase in 
the fraction of unobscured AGNs with increasing X-ray luminosity \citep[e.g.,][]{Treister09b,Trump09}. 
We restrict our sample to $\lx<10^{44}$ \ergs\ to limit the effects of this incompleteness, but we do not reduce the limit further as this would decrease our sample size and dynamic range. 


\section{The prevalence and distribution of AGN accretion activity as a function of stellar mass}
\label{sec:frac}

In this section we investigate how the prevalence and distribution of AGN accretion activity, traced by the number of X-ray-detected AGNs and their X-ray luminosities, depends on the stellar masses of potential host galaxies. 
We consider two broad redshift bins that evenly divide the redshift range probed by PRIMUS, $0.2<z<0.6$ and $0.6<z<1.0$ as shown in Figures \ref{fig:lx_vs_z} and \ref{fig:mass_vs_z}.
Our choice of two redshift bins is a compromise between including enough objects to reveal trends with luminosity or mass, yet avoid serious biases due to possible redshift-dependent completeness or evolutionary effects (evolution of the fraction with redshift is investigated in Section \ref{sec:zevol}). 

Within each redshift bin we subdivide our galaxy sample into bins of different \Mstel;
we then identify X-ray AGNs within each of these stellar mass bins. 
The distribution of X-ray luminosities, \Lx, for objects \emph{within this \Mstel\ bin} is given by $p(\lx \giv \mstel,z)$, the conditional probability density function describing the probability of a galaxy of a given \Mstel\ and $z$ hosting an AGN with luminosity \Lx.
We define $p(\lx \giv \mstel,z)$ as the probability density per \emph{logarithmic} luminosity interval, therefore with units of dex$^{-1}$.
We use the term ``the prevalence of AGN activity"  to refer to the probability of a galaxy hosting an AGN defined by this function, whereas ``the distribution of AGN activity" refers to the distribution of black hole growth traced by the luminosity dependence of this function. The goal of this section is to measure $p(\lx \giv \mstel,z)$ in our two redshift bins.

We can relate $p(\lx \giv \mstel,z)$ to the AGN fraction, \fagn, which we define as the fraction of galaxies of a \emph{given} stellar mass and at a \emph{given} redshift that host an X-ray (non--broad-line) AGN with $\lx>10^{42}$ \ergs, thus
\begin{equation}
f_\mathrm{AGN} (\mstel,z) = \int_{\log \lx=42}^{\infty}  p(\lx \giv \mstel,z) \; \dd \log \lx.
\label{eq:fagnfromplx}
\end{equation}
We note that a basic ``AGN fraction", the fraction of galaxies with an AGN, is a poorly defined and somewhat ad hoc quantity that can depend on the depths of the data and the method used to identify AGNs, in addition to the selection of the parent galaxy sample.
This is certainly true when using X-ray data to identify AGNs; the fraction of galaxies that are X-ray sources is highly dependent on the depths of the X-ray data and the redshifts and magnitude limits of the parent galaxy sample. 
Our quantity of interest, $p(\lx \giv \mstel,z)$, is clearly defined and contains additional information on the distribution as a function of X-ray luminosity.

\subsection{Completeness weighting}

For our initial investigations we bin our sample by redshift, stellar mass, and X-ray luminosity, and directly measure $p(\lx \giv \mstel,z)$ in these bins.
The observed distribution of X-ray luminosities for a given \Mstel\ and $z$ bin will not directly correspond to $p(\lx \giv \mstel,z)$ as the varying sensitivity of our X-ray data prevents us from detecting low-luminosity or high-redshift X-ray AGNs in all galaxies of a given \Mstel\ within our sample. 
It is thus vital to correct for these effects to accurately measure $p(\lx \giv \mstel,z)$ and reveal the relationship between AGN activity and the stellar masses of their host galaxies.
To perform X-ray completeness corrections within each bin we first determine the total number of galaxies, $N_\mathrm{gal}$, within a mass and redshift bin shown in Figure \ref{fig:mass_vs_z}.
We then identify X-ray-detected AGNs within this same mass and redshift bin. 
We take the X-ray luminosity of each source, $L_k$, and calculate $p_\mathrm{det}(L_k\;,\;z_j)$---the probability of being able to detect an AGN with this X-ray luminosity $L_k$ at the redshift $z_j$ of each galaxy \emph{if} the galaxy hosted an AGN of this luminosity.
This is determined by the sensitivity of our X-ray data.
The area of our survey sensitive to a given flux, $A(f_\mathrm{X})$, is described by the X-ray area curves determined in Section \ref{sec:xrayselection}. 
Thus,
\begin{equation}
p_\mathrm{det}(L_k\;,\;z_j)=\frac{A\big(f_\mathrm{X}(L_k,z_j)\big)}{A_\mathrm{total}}
\label{eq:pdet}
\end{equation}
where $A_\mathrm{total}=3.36$ \degsq, the total area covered by both PRIMUS and our X-ray observations, and $f_\mathrm{X}(L_k,z_j)$ is the hard band X-ray flux that would be observed from a source of luminosity $L_k$ at redshift $z_j$. 
We sum these probabilities over all galaxies in our mass and redshift bin to determine a weight for each X-ray source,
\begin{equation}
w_k=\dfrac{N_\mathrm{gal} }{ \sum_{j=1}^{N_\mathrm{gal}}  p_\mathrm{det}(L_k\;,\;z_j) }
\end{equation}
where $w_k$ is the weight for the $k$th X-ray-detected AGN.

Finally, we obtain completeness-corrected estimates of $p(\lx \giv \mstel,z)$ in a range of X-ray luminosity bins by summing these weights within each luminosity bin and dividing by the total number of galaxies (in our given stellar mass and redshift bin). Thus,
\begin{equation}
p(L_m \giv \mstel,z) = \dfrac{ \displaystyle\sum^{ N_m^\mathrm{AGN} }_{k=1}   w_k }
			                						   { N_\mathrm{gal}  \Delta \log L_m}
\end{equation}	
where we sum over all $N_m^\mathrm{AGN}$ AGNs in the $m$th X-ray luminosity bin of width $\Delta \log L_m$. 
We calculate an error on $p(L_m \giv \mstel,z)$  based on the Poisson error in the number of X-ray detected sources in a bin \citep[applying the formulae of ][]{Gehrels86}; we do not consider the contribution of the statistical error in the total number of galaxies in a bin which is minimal due to the very large numbers of sources compared to the X-ray-detected population.
Our completeness corrections make the reasonable assumption that the AGNs (and their host galaxies) that are not X-ray detected due to incompleteness have the same properties as the sources that are detected at the same \Lx, \Mstel, and $z$.

\subsection{Results}

\begin{figure*}
\includegraphics[width=0.49\textwidth]{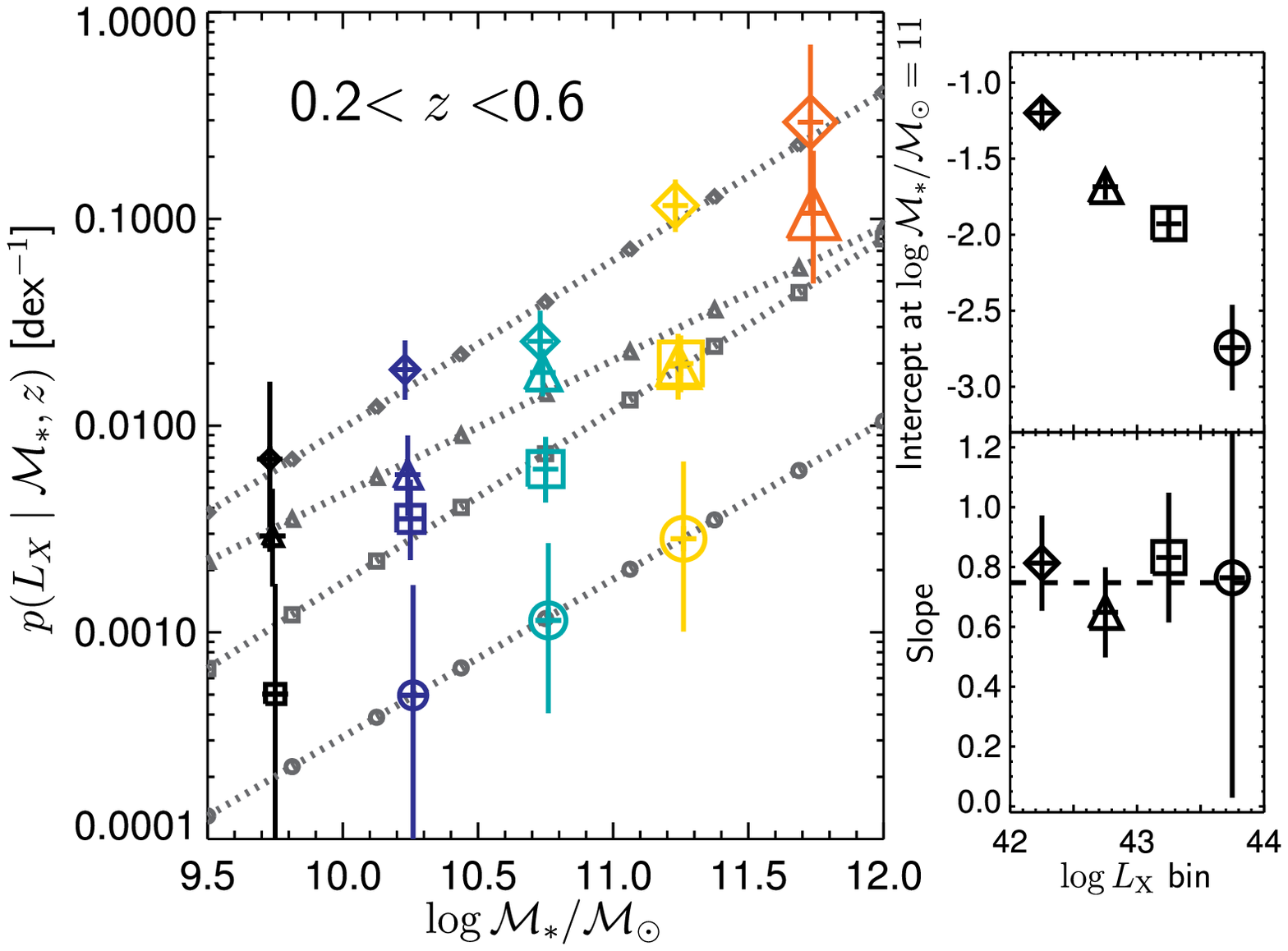}
\hspace{0.02\textwidth}
\includegraphics[width=0.49\textwidth]{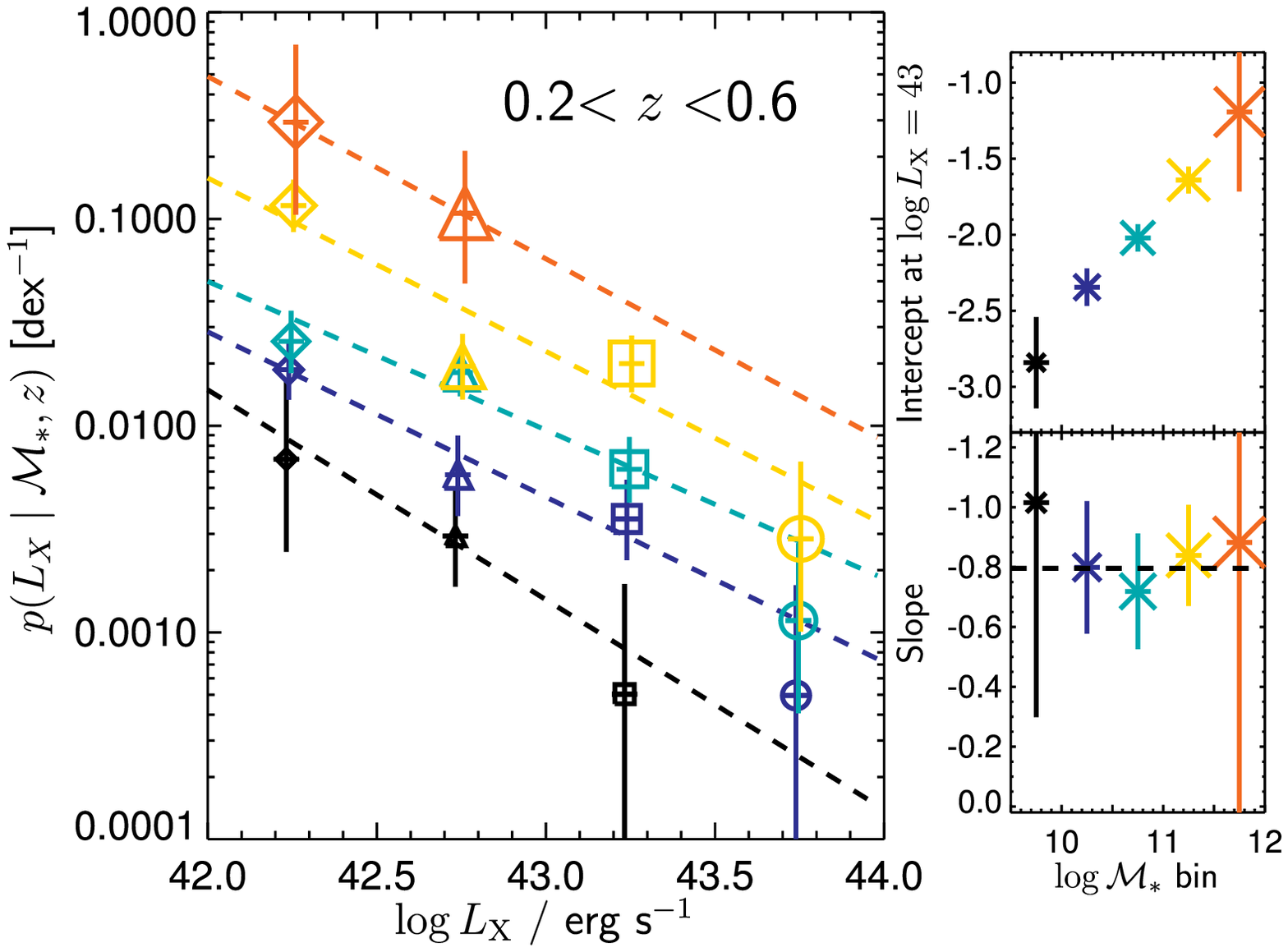}
\vspace{0.3cm}\\
\includegraphics[width=0.49\textwidth]{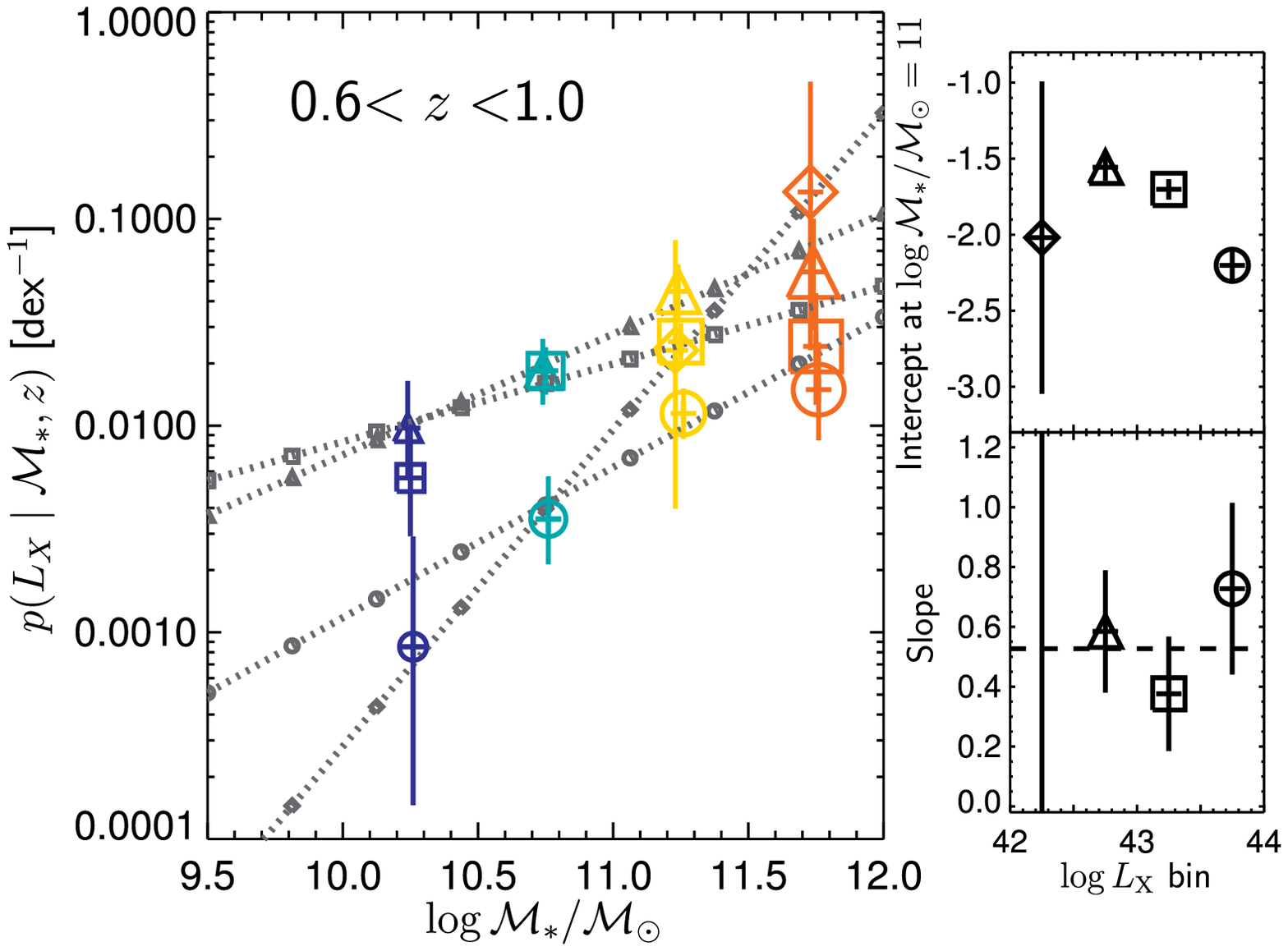}
\hspace{0.02\textwidth}
\includegraphics[width=0.49\textwidth]{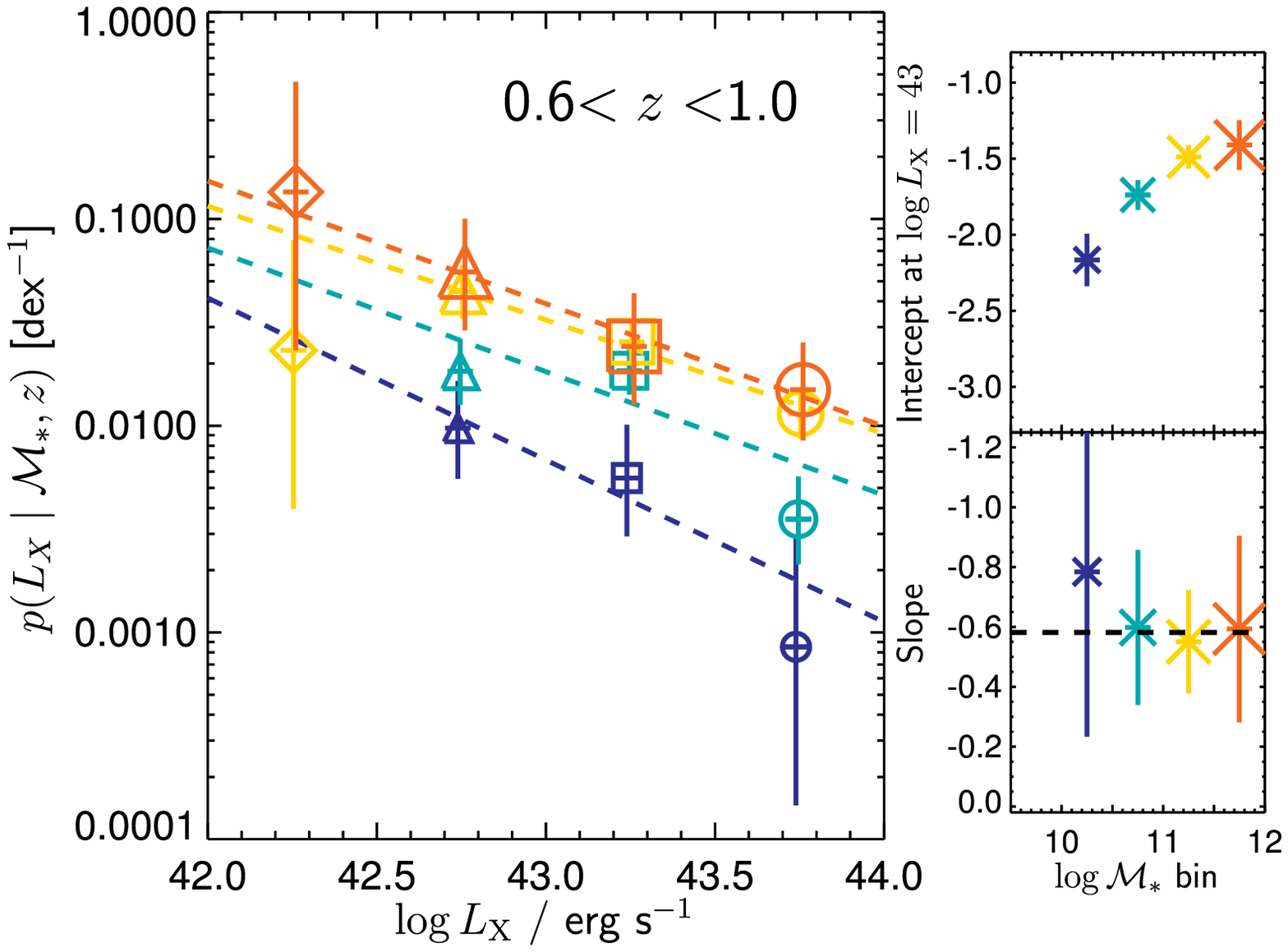}
\caption{
Estimates of $p(\lx \giv \mstel,z)$, the probability density for a galaxy of given stellar mass, \Mstel, and redshift, $z$, to host an AGN of X-ray luminosity, \Lx.
Integrating this function over \Lx\ gives the AGN fraction at a given stellar mass and redshift (see Section \ref{sec:frac}, Equation \ref{eq:fagnfromplx}).
Results are shown as a function of \Mstel\ (left panels) and \Lx\ (right panels) for two redshift ranges: $0.2<z<0.6$ (top) and $0.6<z<1.0$ (bottom). Colors/symbol sizes indicate the \Mstel\ bin and symbol types indicate the \Lx\ bin in both the left and right panels. 
Error bars are based on the Poisson error in the number of X-ray detected AGN within the bin. Lines/small symbols show best fit power-law relations for $p(\lx \giv \mstel,z)$ as a function of \Mstel\ (left) or \Lx (right) for each \Lx\ bin (left, different symbols) or \Mstel\ bin (right, different colors) as defined in Equations \ref{eq:pl_mstel} and \ref{eq:pl_lx} respectively. 
The inset panels show the best-fit slopes and intercepts of these power-law relations. The horizontal black dashed lines show the weighted means of the slopes in the different \Mstel\ and \Lx\ bins. We find that $p(\lx \giv \mstel,z)$ increases at higher \Mstel, but the slope of the relation is consistent in each \Lx\ bin (left panels); $p(\lx \giv \mstel,z)$ also falls with increasing \Lx, but the slope does not depend on the \Mstel\ bin (right panels).
}
\label{fig:fraclxmass}
\end{figure*}

Our estimates of $p(\lx \giv \mstel,z)$ for our two redshift bins are shown in Figure \ref{fig:fraclxmass} where different colors indicate different stellar mass bins and different symbols indicate different X-ray luminosity bins. Two general trends immediately apparent in both our redshift bins are as follows:
\begin{enumerate}
\item The probability of a galaxy hosting an AGN \emph{of a given X-ray luminosity} increases significantly as stellar mass increases.
\item The probability of a galaxy \emph{of a given stellar mass} hosting an AGN decreases as X-ray luminosity increases.
\end{enumerate}
A number of studies have shown that AGNs are found predominately in massive galaxies \citep[e.g.,][]{Best05,Brusa09,Xue10}.
\amend{
We also demonstrate that the probability of hosting an AGN decreases with increasing X-ray luminosity \emph{at all stellar masses}.}
Luminous AGNs are much rarer than their low-to-moderate-luminosity equivalents, as shown by the AGN luminosity function for the overall population
\citep{Ueda03,Aird10}, thus the overall trend of fewer luminous AGNs is expected.
However, more luminous AGNs may be expected to be associated with the most massive galaxies, while low-mass galaxies might be expected to host the low-luminosity AGN population. 
Our results show that this is not the case. 
Indeed, for a sample of galaxies of \emph{any} given mass, a higher fraction will be found to host low-luminosity AGNs. The X-ray luminosity distribution of AGNs does not appear to depend strongly on the host galaxy mass; equivalently the probability of hosting an AGN as a function of host galaxy mass appears to be independent of the X-ray luminosity. 

To further investigate these trends we perform $\chi^2$ fits to our AGN fractions using our binned estimates and Poisson errors. 
Motivated by the distributions seen in Figure \ref{fig:fraclxmass}, which show roughly linear trends in log--log space, we assume simple power-law relations for $p(\lx \giv \mstel,z)$ as a function of stellar mass at a fixed X-ray luminosity,
\begin{equation}
\log\left[  p( \lx \giv \mstel,z) \right]= a + b\log\left[ \frac{ \mathcal{M}_* }{ 10^{11} M_\sun} \right],
\label{eq:pl_mstel}
\end{equation}
where $a$ and $b$ are the intercept and power-law slope of the relation. The best-fit values for each $L_\mathrm{X}$ bin are shown in the small insets on the left side of Figure \ref{fig:fraclxmass}. The fits are consistent with a slope that does not change with X-ray luminosity, in agreement with our basic observation that the variation of $p(\lx \giv \mstel,z)$ as a function of \Mstel\ is independent of $L_\mathrm{X}$.
We also fit power-law relations for $p(\lx \giv \mstel,z)$ as a function of X-ray luminosity for a fixed stellar mass,
\begin{equation}
\log\left[  p(\lx \giv \mstel,z) \right]= c + d\log\left[ \frac{ L_\mathrm{X} }{ 10^{43} \;\mathrm{erg\; s^{-1}} }\right],
\label{eq:pl_lx}
\end{equation}
with intercept $c$ and slope $d$.
The best fit parameters for each stellar mass bin are shown in the inset panels on the right side of Figure \ref{fig:fraclxmass}. Again, the fits are consistent with the distribution of AGN luminosities being independent of stellar mass. 

We note that the same trends are seen in both redshift bins. \amend{Given large errors,} the slopes of our fitted power laws are consistent between our redshift bins. There is also weak evidence that the intercepts, \moreamend{shown in the inset panels}, are generally larger at higher redshift, indicating evolution with redshift in the overall fraction. This is investigated in more detail in the next section. 


\section{Redshift evolution}
\label{sec:zevol}

To investigate any evolution of $p(\lx \giv \mstel,z)$ with redshift we must divide our sample into finer redshift bins. This reduces the number of sources in each of our stellar mass and luminosity bins.
The $\chi^2$ fitting approach requires fairly heavy binning and will be sensitive to the chosen bin sizes.
If the bins are too large then the actual trends in the data may be hidden, and the individual binned measurements may be systematically biased by large sensitivity variations within a bin; 
if the bins are too small then they will contain too few objects and $\chi^2$ statistics cannot be used.
Binning also means that we do not utilize all the available information: the actual measurements of \Mstel\ and redshift for each individual galaxy and the \Lx\ measurements for each X-ray detected AGN.
We therefore develop a maximum likelihood fitting approach to determine the dependence of $p(\lx \giv \mstel,z)$ on \Mstel\ and \Lx\ in a number of smaller redshift bins.
In Section \ref{sec:inc_zevol} we directly incorporate redshift evolution into this scheme, removing the need to bin by redshift.

\subsection{Maximum likelihood fitting}
\label{sec:mlfit}

The results of the previous section imply that $p(\lx \giv \mstel,z)$ at a fixed redshift may be written as a separable function of stellar mass and X-ray luminosity of the form
\begin{equation}
p(\lx \giv \mstel,z) \;\dd\log\lx =
	K \left(\frac{\mstel}{\mathcal{M}_0}\right)^{\gamma_\mathcal{M}}
	    \left(\frac{L_\mathrm{X}}{L_\mathrm{0}}\right)^{\gamma_L} \dd \log\lx,
\label{eq:fagnfunc}	 
\end{equation}
where $\mathcal{M}_0$ and $L_0$ are arbitrary scaling factors, $K$ is an overall normalization and $\gamma_\mathcal{M}$ and $\gamma_L$ are power-law slopes. 
We adopt this functional form to perform an unbinned extended maximum-likelihood fit to our data, and measure values of our three free parameters ($K$, $\gamma_\mathcal{M}$ and $\gamma_L$) in each of our redshift bins.
The extended maximum-likelihood fit allows us to directly constrain the normalization, $K$, along with the other model parameters ($\gamma_L$, $\gamma_\mathcal{M}$) and estimate robust confidence intervals on all three parameters. To find the best values we maximize the log-likelihood function,
\begin{equation}
\ln \mathcal{L}= -\mathcal{N} + \sum_{k=1}^{N_i^\mathrm{AGN}} \ln p_k
\label{eq:loglik}
\end{equation}
where the summation is performed over all $N_i^\mathrm{AGN}$ X-ray-detected AGNs in the $i$th redshift bin, and $p_k$ is the probability that a galaxy of stellar mass $\mathcal{M}_k$ will host an AGN of X-ray luminosity $L_k$.
This is directly related to $p(\lx \giv \mstel,z)$, thus
\begin{equation}
p_k=p(L_k \giv \mathcal{M}_k, z_k)=K  \left(\dfrac{\mathcal{M}_k}{\mathcal{M}_0}\right)^{\gamma_\mathcal{M}}
	 	    \left(\dfrac{L_k}{L_\mathrm{0}}\right)^{\gamma_L},
\label{eq:pk}		  
\end{equation}
where we assume no dependence on $z$ within our fine redshift bin.
The \emph{expected} number of X-ray AGNs, $\mathcal{N}$, for a given set of model parameters is found by summing the probability of \emph{observing} an AGN of any luminosity $\log L_\mathrm{X}>42$ over all $N_i^\mathrm{gal}$ galaxies in our redshift bin. 
The probability of observing an AGN is the combination of the intrinsic probability density function, $p(\lx \giv \mstel,z)$, which gives the probability that a galaxy hosts an AGN of luminosity \Lx, and the probability of actually being able to detect such a source, $p_\mathrm{det}(\lx,z)$, defined in Equation \ref{eq:pdet}. Thus,
\begin{eqnarray}
	\mathcal{N}&=& \displaystyle \sum^{N^\mathrm{gal}_i}_{j=1}
					 \int^{\infty}_{42} p(\lx \giv \mathcal{M}_j, z_j) \; p_\mathrm{det}(\lx,z_j) \; \dd\log\lx \\
		&=& \displaystyle\sum^{N^\mathrm{gal}_i}_{j=1} 
			K  \left(\frac{\mathcal{M}_j}{\mathcal{M}_0}\right)^{\gamma_\mathcal{M}}
			  \int^{\infty}_{42} 
				     \left(\frac{L_\mathrm{X}}{L_\mathrm{0}}\right)^{\gamma_L} 
					p_\mathrm{det}(\lx,z_j) 	\; \dd \log\lx . \nonumber
\label{eq:npred}
\end{eqnarray}
The X-ray completeness is accounted for by the $p_\mathrm{det}(\lx,z_j)$ term; this reduces the expected number of AGN at low luminosities and high redshifts due to the X-ray sensitivity limits of our survey. We set $\log L_\mathrm{min}=42.0$ as the cutoff luminosity above which an X-ray source is associated with AGN activity. 
We estimate equivalent $1\sigma$ errors on the model parameters from the maximum projection of the $\Delta \mathcal{S}=1.0$ likelihood surface for each parameter, where $\mathcal{S}=-2 \ln \mathcal{L}$.

\subsection{Results in finer redshift bins}

\begin{figure}
\begin{center}
\includegraphics[width=0.4\textwidth]{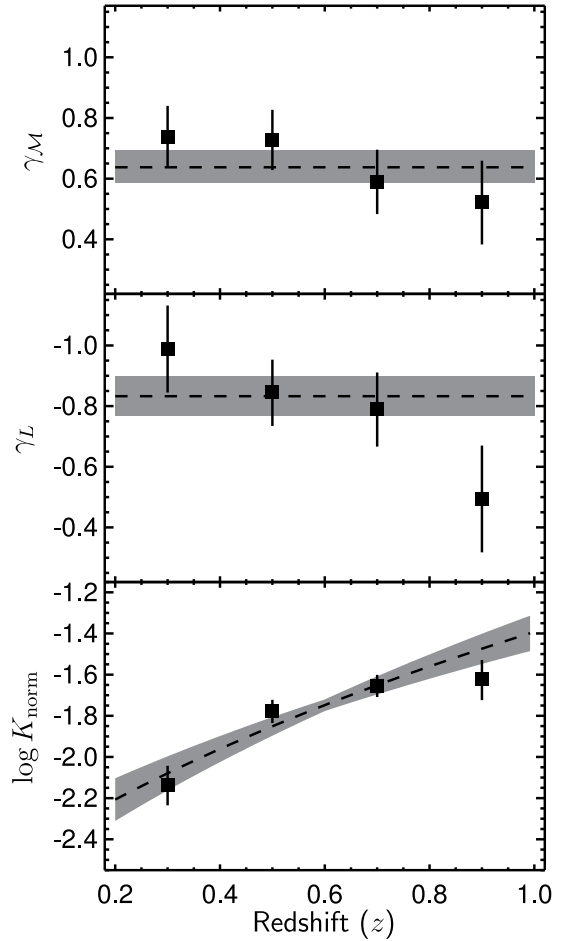}
\end{center}
\caption{
Best-fit parameters from the maximum-likelihood fitting for $p(\lx \giv \mstel,z)$. 
Squares show the best fit parameters (and 1$\sigma$ equivalent errors) fitting in four redshift bins, assuming that $p(\lx \giv \mstel,z)$ is constant over the redshift bin. 
The dashed lines are the results fitting over the entire redshift range, allowing for redshift evolution as described by Equation \ref{eq:zevol}: the dashed lines in the top two panels show the best estimates of $\gamma_\mathrm{M}$ and $\gamma_L$ for the entire redshift range (gray regions indicate 1$\sigma$ uncertainty); in the bottom panel the dashed line indicates the equivalent of $K$ for the model with redshift evolution (see Equation \ref{eq:kequiv}). We find that $p(\lx \giv \mstel,z)$ is described by independent power-law functions of \Mstel\ and \Lx\ that do not change with redshift; there is however significant evolution in the normalization that follows a power-law dependence on $(1+z)$.
}
\label{fig:parsvsz}
\end{figure}

We have used the maximum likelihood fitting approach described above to constrain the functional form of $p(\lx \giv \mstel,z)$ in four redshift bins, evenly spaced across the $0.2<z<1.0$ range probed by our sample. We fix the scaling factors at $\mathcal{M}_0=10^{11} M_\sun$ and $L_0=10^{43}$ \ergs. We note that our current scheme makes the simplifying assumption that $p(\lx \giv \mstel,z)$ does not strongly depend on redshift within the range of the redshift bin. 
The best fit parameters are shown by the points and error bars in Figure \ref{fig:parsvsz}. We find that the power-law slopes, $\gamma_\mathcal{M}$ and $\gamma_L$, do not change significantly with redshift.
We do, however, find that the normalization increases significantly with increasing redshift. Thus, the fraction of galaxies \emph{of a given stellar mass} hosting AGN \emph{of a given X-ray luminosity} increases substantially with redshift, indicating a higher prevalence of AGN accretion activity at earlier times.

\subsection{Incorporating redshift evolution}
\label{sec:inc_zevol}

We modify our maximum-likelihood estimator to include redshift evolution, allowing us to directly constrain the form of the observed redshift evolution and remove redshift binning from our analysis. 
We assume that the redshift evolution is independent of X-ray luminosity or stellar mass and has a simple power-law form in $(1+z)$, and thus we modify Equation \ref{eq:fagnfunc} to 
\begin{eqnarray}
p(\lx & &\giv \mstel,z) \;\dd \log \lx = \label{eq:zevol} \\
	& &A	 \left(\frac{\mathcal{M}}{\mathcal{M}_0}\right)^{\gamma_\mathcal{M}}
	 \left(\frac{L_\mathrm{X}}{L_\mathrm{0}}\right)^{\gamma_L}
	 \left(\frac{1+z}{1+z_0}\right)^{\gamma_z}	\dd\log\lx  \nonumber
\end{eqnarray}
where we set the scaling factor $z_0=0.6$, and $A$ is an overall normalization. 
We adapt Equations \ref{eq:pk} and \ref{eq:npred} to incorporate this functional form and perform our maximum-likelihood fitting over our entire $0.2<z<1.0$ redshift range. 
Our best-fit parameters and their uncertainties are given in Table \ref{tab:fitpars}. We also show the best-fit values and uncertainties of the $\gamma_\mathcal{M}$ and $\gamma_L$ slopes in the top two panels of Figure \ref{fig:parsvsz} by the dashed lines and gray shaded regions respectively. 
In the bottom panel of Figure \ref{fig:parsvsz} we show the equivalent normalization, 
\begin{equation}
K=A [ (1+z)/(1+z_0) ]^{\gamma_z},
\label{eq:kequiv}
\end{equation}
based on our best fit evolutionary model.
We find evidence for very strong evolution in the AGN fraction with redshift of the form $p(\lx \giv \mstel,z)\propto (1+z)^{4}$.

\begin{table}
\caption{Best-fit Parameters from Maximum-likelihood Fitting with Redshift Evolution}
\begin{center}
\begin{tabular}{c c}
\vspace{-5mm}\cr
\hline
\hline
\vspace{-2 mm} \cr
Parameter & Value\\
\vspace{-2 mm} \cr
\hline 
\vspace{-2 mm} \cr
$\log_{10} A$ &  $-$\amend{1.75}$^{+ 0.03}_{- 0.03}$ \\
$\gamma_\mathcal{M}$ &  \phs\amend{0.64}$^{+ 0.06}_{-0.05}$ \\
$\gamma_L$ &  $-$\amend{0.83}$^{+ 0.06}_{- 0.07}$ \\
$\gamma_z$ &  \phs\amend{3.67}$^{+ 0.60}_{- 0.61}$ \\
\vspace{-2 mm} \cr    
\hline
\end{tabular}
\end{center}
\label{tab:fitpars}
\end{table}

\begin{figure*}
\plottwo{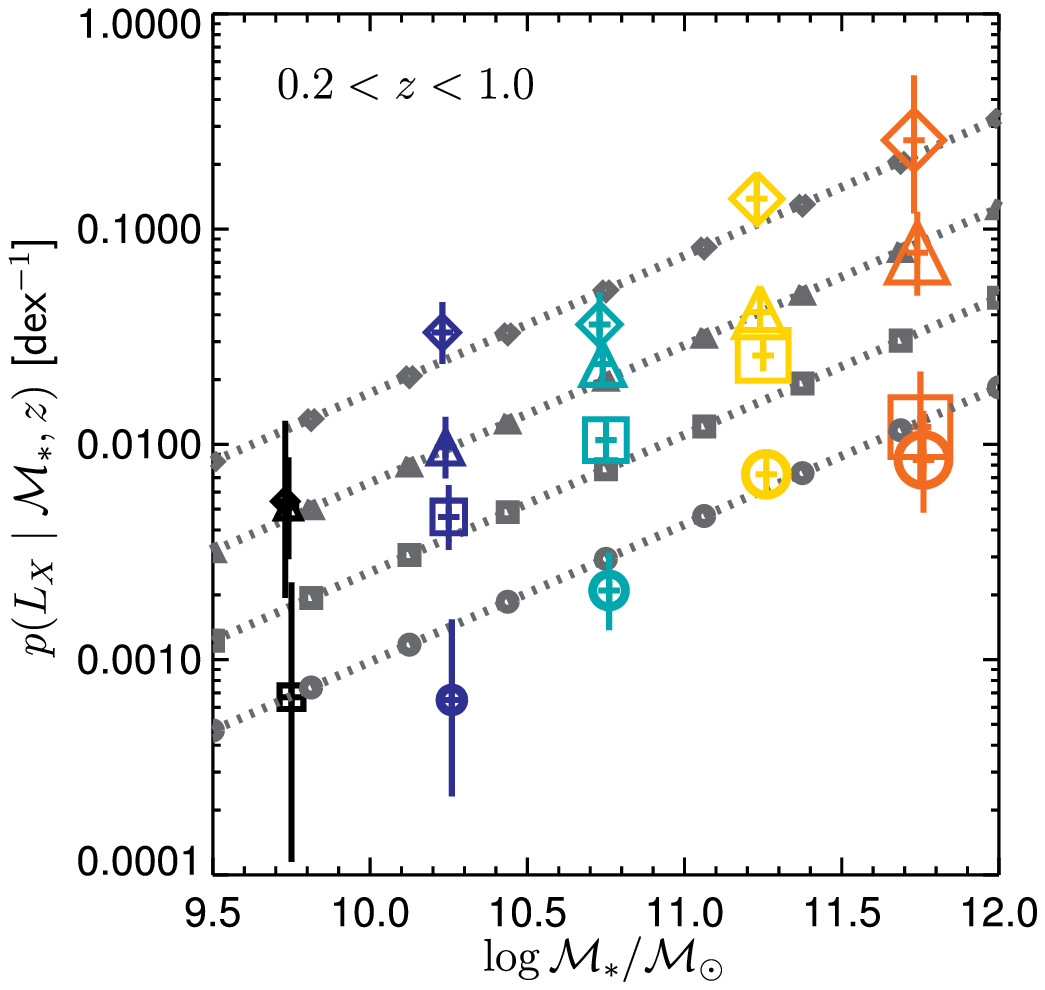}{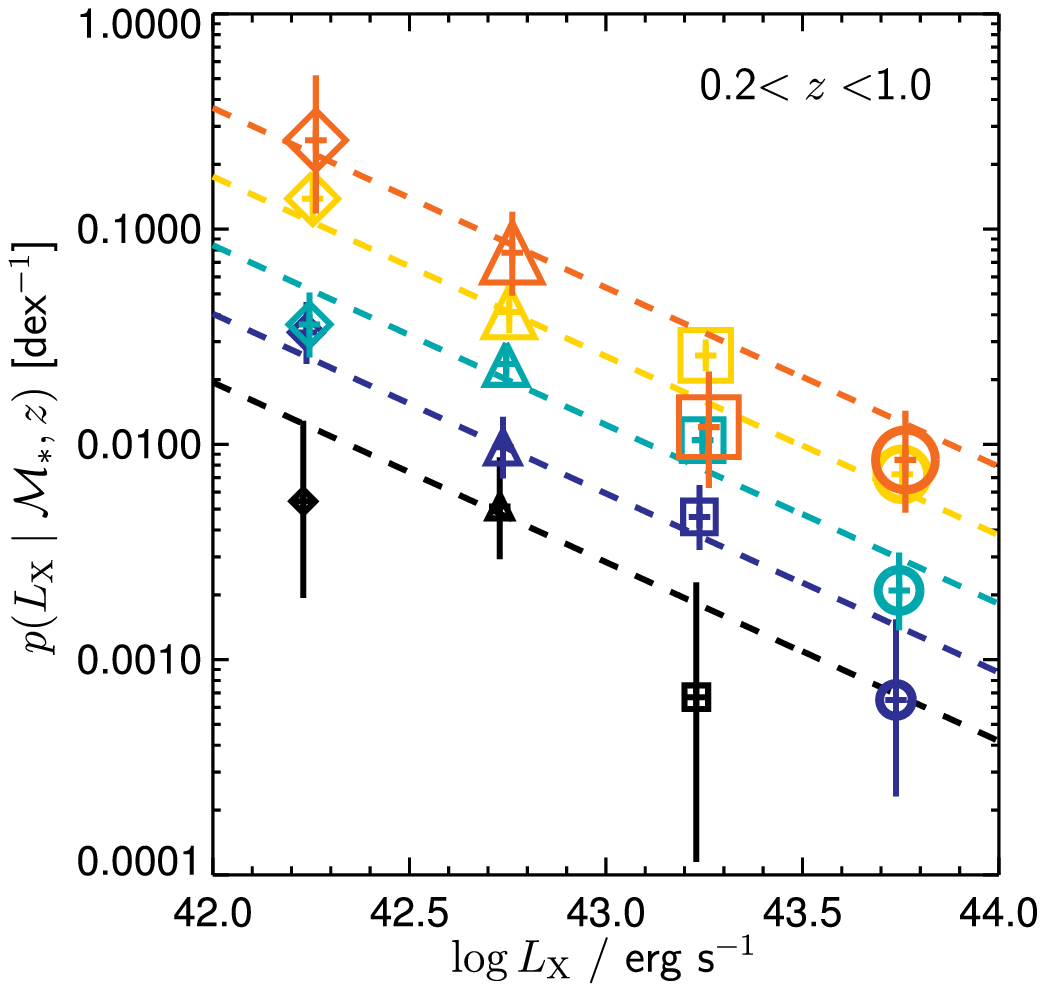}
\caption{
Estimates of $p(\lx \giv \mstel,z)$ as a function of \Mstel\ (left) and \Lx\ (right) based on our maximum-likelihood fitting results. We combine data over our entire redshift range ($0.2<z<1.0$). 
The small gray symbols and lines (left) and colored lines (right) show our best fit model from the unbinned maximum-likelihood fitting (Equation \ref{eq:zevol}, Table \ref{tab:fitpars}) evaluated at the center of our redshift range ($z=0.6$). 
Binned data points are estimated using the $N_\mathrm{obs}/N_\mathrm{mdl}$ method (Section \ref{sec:nobsnmdl}), which compares the observed number of AGNs with that predicted within an \Mstel\ and \Lx\ bin using our best-fit model, given our observed parent galaxy sample. 
This allows us to correct for strong redshift, \Mstel\ and \Lx\ dependent effects on the binned estimates, and directly compare deviations of the data from the model.
Colors/symbol sizes and symbol types correspond to the \Mstel\ and \Lx\ bins, respectively (as in Figure \ref{fig:fraclxmass}). 
Error bars are based on the Poisson error in the number of X-ray detected AGNs within the bin.
}
\label{fig:bestfrac}
\end{figure*}

\label{sec:nobsnmdl}

To aid visualization of our data and comparisons with our models we implement the $N_\mathrm{obs}/N_\mathrm{mdl}$ method \citep{Miyaji01}. 
This method accounts for significant variations of both our model function and the X-ray sensitivity over a bin.
This is especially useful given our relatively small sample of X-ray-detected AGNs, the large variation of the X-ray sensitivity as a function of \Lx\ and redshift, and the strong redshift dependence we have found. This allows us to accurately combine data over large bins and identify true deviations of the data from the model.

The $N_\mathrm{obs}/N_\mathrm{mdl}$ method compares the observed number of sources in a bin, $N_\mathrm{obs}$, with the predicted number based on a model fit. The binned estimate is then given by the model evaluated at the center of the bin, scaled by the ratio of $N_\mathrm{obs}/N_\mathrm{mdl}$. 
For our purposes, $N_\mathrm{mdl}$ is the predicted number of galaxies within a given redshift and mass bin that we expect to detect as an X-ray AGN over a given range of luminosities, for a particular set of model parameters. Thus,
\begin{eqnarray}
& N_\mathrm{mdl}(\mathcal{M}_i,z_i,L_m) = 
 \displaystyle\sum_{k=1}^{N_i^\mathrm{gal}} & \Bigg[ A \left(\dfrac{1+z_i}{1+z_0}\right)^{\gamma_z}
 \left(\dfrac{\mathcal{M}_i}{\mathcal{M}_0}\right)^{\gamma_\mathcal{M}} \nonumber\\ 
&& 
 \displaystyle \int_{\log L_m^\mathrm{lo}}^{\log L_m^\mathrm{hi}} \left(\frac{L_\mathrm{X}}{L_\mathrm{0}}\right)^{\gamma_L} \; \dd \log L_\mathrm{X} \Bigg] \nonumber\\
\end{eqnarray}
where the summation is over all $N_i^\mathrm{gal}$ galaxies in the $i$th redshift and stellar mass bin, and $L_m^\mathrm{hi}$ and
$L_m^\mathrm{lo}$ are the upper and lower limits of the $m$th X-ray luminosity bin. 

In Figure \ref{fig:bestfrac} we show $p(\lx \giv \mstel,z)$ as a function of stellar mass and luminosity over the full redshift range ($0.2<z<1.0$) with $N_\mathrm{obs}/N_\mathrm{mdl}$ binned estimates. The model lines are evaluated at $z=0.6$, but our binned estimates use data over the full redshift range and are corrected for the strong redshift evolution we find in the AGN fraction. 
The binned estimates show that our observed data are well described by our model.


\section{Eddington ratio distribution}
\label{sec:eddratio}

In this section we investigate how the fraction of galaxies hosting an obscured X-ray AGN depends on the Eddington ratio: the ratio of the bolometric luminosity to the Eddington limit, $L_\mathrm{Edd}$, where radiation pressure balances the infall of material under gravity (assuming spherical symmetry and hydrostatic equilibrium).
The Eddington ratio tracks the rate at which matter is accreted by the SMBH and thus relates to the mode of accretion and the processes that fuel the AGN \citep[e.g.,][]{Kauffmann09,Trump11}.
The Eddington ratio could be the principal factor that determines the prevalence of AGN activity in potential host galaxies.
It is thus vital to accurately measure the distribution of Eddington ratios, account for possible selection effects that will bias against sources accreting at the lowest Eddington ratios, and uncover the relationship between AGN activity and the properties and evolution of their host galaxies. 

\subsection{Indications of an Eddington ratio dependence}

The results of Section \ref{sec:zevol} indicate that the prevalence and distribution of AGN activity is related to the Eddington ratio. 
We find that $p(\lx \giv \mstel,z)$ increases with \Mstel\ at a fixed \emph{X-ray luminosity}. 
However, the mass of the central SMBH that powers an AGN is correlated with the mass of a galaxy's central stellar bulge, as shown by the $\mbh-\sigma$ relation \citep{Ferrarese00,Gebhardt00} as well as more direct probes \citep[e.g.,][]{Haring04}.
A higher stellar mass galaxy is therefore expected to host a higher mass SMBH, and thus the same X-ray luminosity corresponds to a much lower Eddington ratio. 
Therefore, the rise of $p(\lx \giv \mstel,z)$ with \Mstel\ may simply reflect the higher luminosities of low Eddington ratio sources in galaxies with higher stellar masses. 
We also find roughly similar slopes for the dependence of $p(\lx \giv \mstel,z)$ on \Mstel\ and \Lx\, such that $\gamma_L\approx-\gamma_\mathcal{M}$ to within $\sim 2\sigma$.
This indicates that $p(\lx \giv \mstel,z)$ may be dependent on Eddington ratio, $\lambda_\mathrm{Edd}$, and redshift only, assuming a single bolometric correction such that \Lx$\propto L_\mathrm{bol}$ and a direct proportionality between the stellar mass of a galaxy and its central SMBH, \Mstel$\propto \mbh$. Thus we can re-write Equation \ref{eq:zevol} as 
\begin{equation}
p(\lamedd \giv \mstel,z) \; \dd \log \lambda_\mathrm{Edd}
	= A \; \lamedd^{\gamma_E} \left(\frac{1+z}{1+z_0}\right)^{\gamma_z}  \dd \log \lamedd,
\label{eq:fedd}
\end{equation}
where $\gamma_E\approx \gamma_L \approx - \gamma_\mathcal{M} \approx -0.7$. 

\subsection{Bolometric corrections}

The assumption of a single bolometric correction may be an over-simplification. 
Both \citet{Marconi04} and \citet{Hopkins07b} have shown that the 2--10 keV X-ray bolometric correction, $k_\mathrm{2-10 keV}$, is luminosity dependent with an approximately power-law form, based on composite SEDs and the observed distribution of $\alpha_\mathrm{OX}$, the X-ray to optical spectral index \citep{Zamorani81}.
A number of recent studies, however, have used observed multiband SEDs of unobscured AGNs to accurately measure bolometric luminosities and propose that $k_\mathrm{2-10 keV}$ may be more closely correlated with the Eddington ratio, albeit with significant scatter \citep{Vasudevan07,Vasudevan09,Lusso10}.
These studies focused on relatively luminous, unobscured AGNs, that are mostly accreting at high Eddington ratios ($\lamedd\gtrsim0.1$), and the correlation does not appear to hold at lower \Lamedd.
We therefore adopt the luminosity-dependent $k_\mathrm{2-10keV}$ from \citet{Hopkins07b}.
Our bolometric corrections vary from $\sim15-50$, spanning the range of observed values seen in \citet{Vasudevan07} for $\lamedd\lesssim0.1$,
We do not expect uncertainty in the bolometric correction to significantly affect our results given the wide range of Eddington ratios that we probe ($
\sim$four orders of magnitude).

We replace \Lx\ with the bolometric luminosity, $L_\mathrm{bol}$, in Equation \ref{eq:zevol}, and then repeat our maximum-likelihood fitting for the entire galaxy sample. 
Using this bolometric correction we find a slightly flatter slope for the dependence on (bolometric) luminosity, $\gamma_L=-0.70\pm0.05$, which is in good agreement with the inverse of the mass-dependent slope, $\gamma_\mathcal{M}=0.64\pm0.05$, consistent with the prevalence of AGN activity being primarily  determined by the Eddington ratio.

\subsection{Maximum-likelihood fitting of a universal Eddington ratio distribution}

We repeat our maximum-likelihood fitting for our entire galaxy sample using the form given in Equation \ref{eq:fedd}, directly fitting a single power-law slope, $\gamma_E$, for the Eddington ratio. 
We assume a constant scaling between the black hole mass and the host galaxy stellar mass, $\mbh\approx0.002\mstel$ \citep[where we have assumed $\mstel\approx \mathcal{M}_\mathrm{bulge}$]{Marconi03}.
We use this relation to calculate Eddington ratios,
\begin{equation}
\lamedd=\frac{L_\mathrm{bol}}{L_\mathrm{Edd}},
\end{equation}
where the Eddington limit, $L_\mathrm{Edd}$, in units of \ergs\ is given by
\begin{eqnarray}
L_\mathrm{Edd}&=&1.3 \times 10^{38} \;\dfrac{\mbh}{\msun} \nonumber\\
	&=&1.3 \times 10^{38} \;\dfrac{0.002\mstel}{\msun}.
\end{eqnarray}

\begin{figure*}
\includegraphics[width=\textwidth]{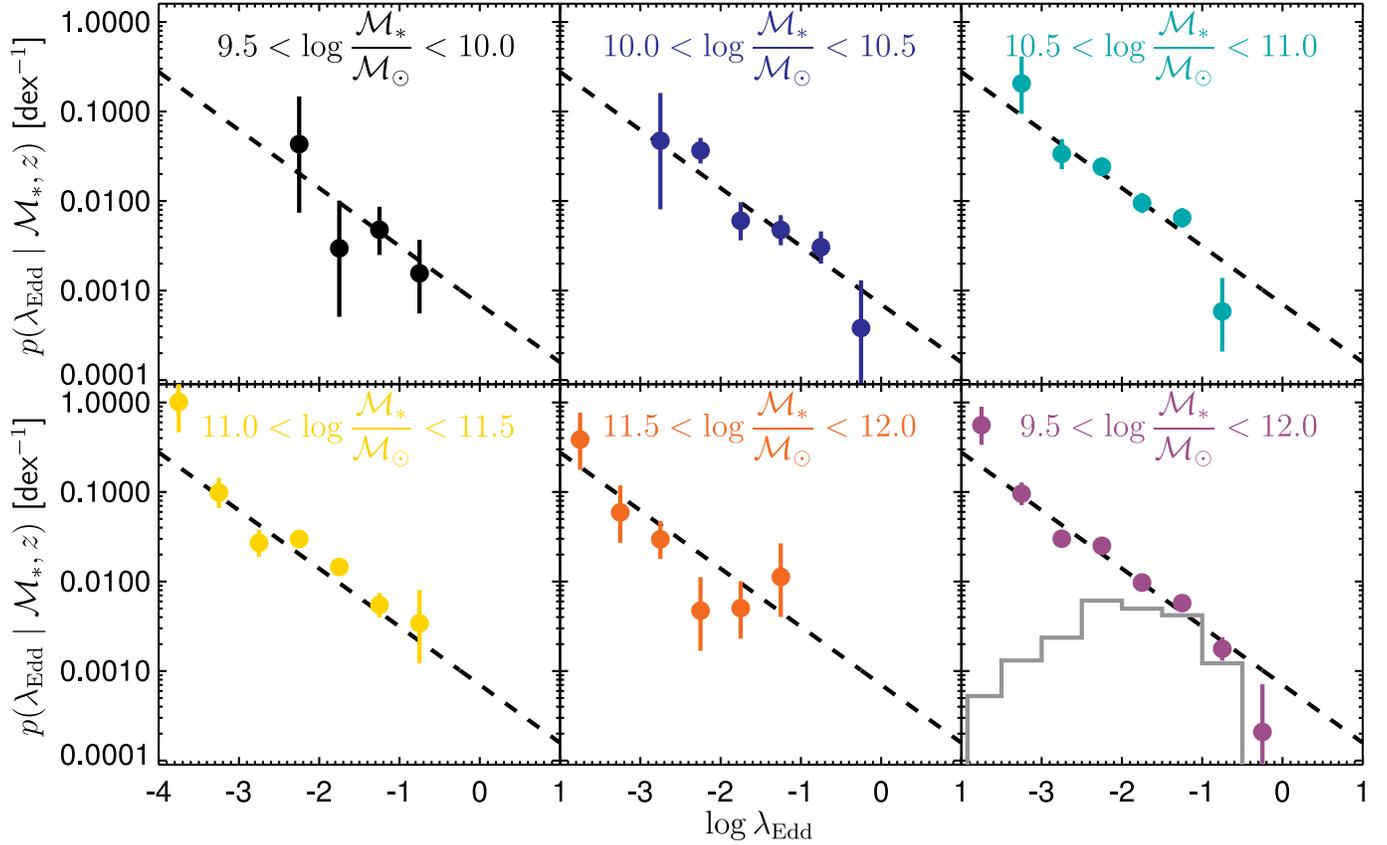}
\caption{
Best-fit model of $p(\lamedd \giv \mstel,z)$ evaluated at $z=0.6$ (dashed lines, which are the same in every panel) and compared to the data for $0.2<z<1.0$ in our various stellar mass bins using the $N_\mathrm{obs}/N_\mathrm{mdl}$ method (points and error bars, see Section \ref{sec:nobsnmdl}). 
The final panel combines data over our entire stellar mass range. 
Our data are consistent with the same distribution of Eddington ratios at every stellar mass, indicating that the probability of a galaxy hosting an AGN of a given \Lamedd\ does \emph{not} depend on stellar mass. 
In the final panel we also show the observed distribution of Eddington ratios (light gray histogram) \emph{without} correcting for the X-ray sensitivity effects or the redshift evolution of $p(\lamedd \giv \mstel,z)$. 
The observed distribution has a roughly lognormal shape with a peak at $\lamedd\approx0.01$ rather than the true power-law shape that we recover.
}
\label{fig:leddfrac_massbin}
\end{figure*}

\begin{figure}
\begin{center}
\includegraphics[width=0.45\textwidth]{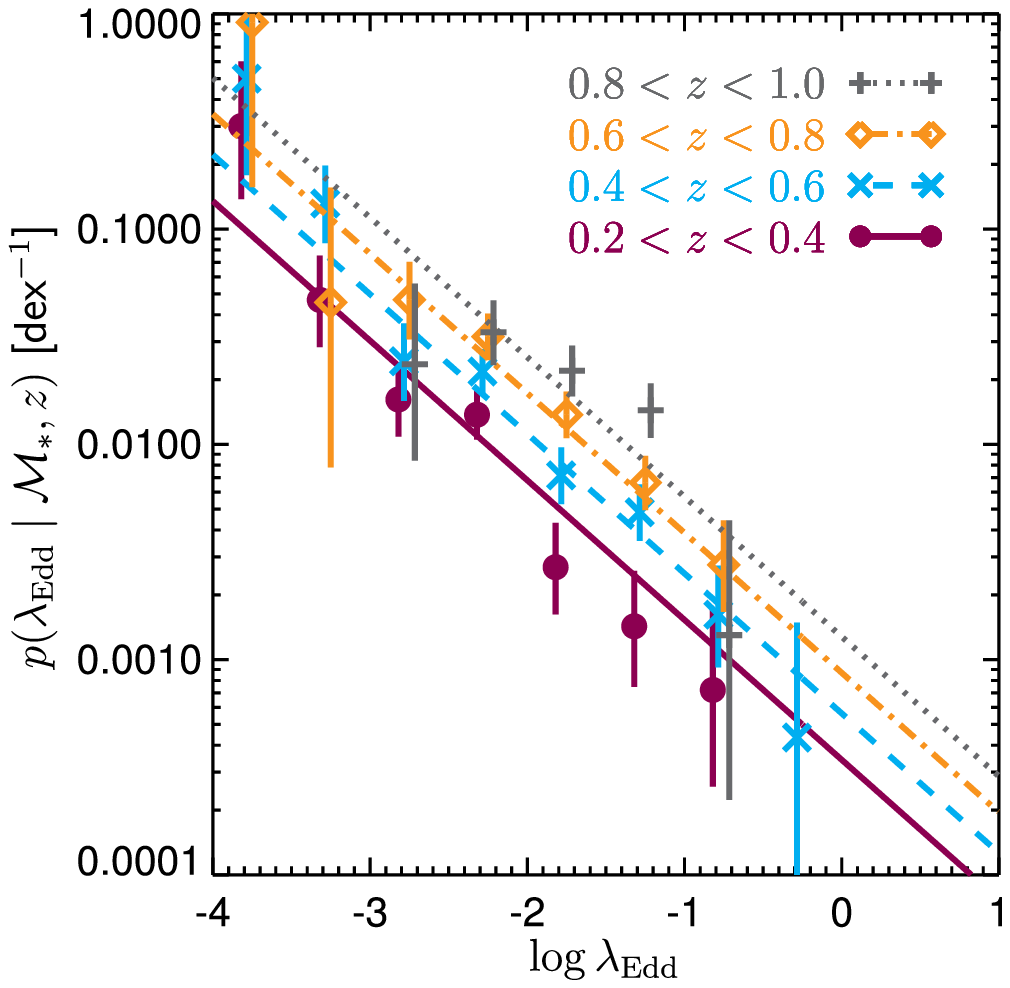}
\end{center}
\caption{
Best-fit model of $p(\lamedd \giv \mstel,z)$ evaluated at the centers of the four indicated redshift bins (lines)
and compared to the data over the given redshift bin and our full range of stellar masses,
$9.5<\log \mstel/\msun<12.0$ using the $N_\mathrm{obs}/N_\mathrm{mdl}$ method (points and error bars, slightly offset from the center of the \Lamedd\ bin to aid comparison). 
We find significant evidence for a strong evolution with redshift 
of the form $p(\lamedd \giv \mstel,z)\propto (1+z)^{3.8}$ 
(given our assumptions for the bolometric luminosity corrections and scaling of \Mstel\ to \Mbh).
However, with our data we cannot distinguish between an overall increase in the density of black holes that are actively accreting at a given \Lamedd\ with increasing redshift, or a shift of the entire population to higher \Lamedd\ (see further discussion in Section \ref{sec:agntriggering}). 
}
\label{fig:leddfrac_zbin}
\end{figure}

The results of our maximum-likelihood fitting for our entire sample with an Eddington ratio and redshift dependence only are given in Table \ref{tab:eddmlfit}. 
In Figure \ref{fig:leddfrac_massbin}, we show $p(\lamedd \giv \mstel,z)$ as a function of \Lamedd\ for galaxies in different stellar mass bins; we combine our entire redshift range and use our $N_\mathrm{obs}/N_\mathrm{mdl}$ approach described in Section \ref{sec:nobsnmdl} to account for the strong redshift evolution. Our results show that the distribution of Eddington ratios does \emph{not} depend on the stellar mass of the galaxies; all of the points are consistent with the same global relation. For a sample of galaxies of any given mass we expect to find the same distribution of Eddington ratios. 
In the final panel of Figure \ref{fig:leddfrac_massbin} we combine data over our full range of stellar masses and redshifts, showing the excellent agreement with our best-fit relation for the \Lamedd\ dependence.
For comparison, we also show the observed distribution of Eddington ratios for our hard X-ray AGN sample \emph{without} accounting for the X-ray sensitivity effects or redshift evolution of $p(\lamedd \giv \mstel,z)$ (gray histogram).
This has a roughly lognormal shape that peaks at $\lamedd\approx 0.01$ rather than the power-law shape that we recover.
In Figure \ref{fig:leddfrac_zbin} we show $p(\lamedd \giv \mstel,z)$ for our full stellar mass range but in different redshift bins; we find a significant increase in the fraction of galaxies that host AGNs with redshift, although the distribution of Eddington ratios follows the same power-law form. 

Our results indicate that AGNs are \emph{not} predominantly in massive galaxies when considered in terms of Eddington ratio.
Above a given \emph{X-ray luminosity}, a higher fraction of AGNs will be found in more massive galaxies, but this is a selection effect as low Eddington ratio AGNs in high stellar mass galaxies will have higher X-ray luminosities.
We find that the Eddington ratio distribution is a universal function with a power-law shape of slope $-0.65$.

\subsection{Specific accretion rates}
\label{sec:specaccrate}

The Eddington ratio distributions presented in this section assume that the stellar mass of a galaxy is directly proportional to the mass of the central SMBH. 
However, the $\mbh-\sigma$ relation implies a correlation between the mass of the stellar \emph{bulge} and the SMBH.
The fraction of the total stellar mass contained in the bulge will depend on the galaxy morphology, which in turn is related to both the color 
and stellar mass of the host galaxy \citep[see][]{Blanton09}, introducing significant scatter to our \Mbh\ estimates.
It is also unclear if the $\mbh-\sigma$ relation evolves with redshift or exists at all outside the local universe \citep[e.g.,][]{Shields03,Merloni04,Croton06b,Peng06,Woo06,Salviander07,Woo08,Merloni10,Shen10}. 
In fact, \citet{Jahnke09} propose that while the $\mbh-\mathcal{M}_\mathrm{bulge}$ relationship may evolve with redshift, there is little evolution in the correlation between \Mbh\ and the \emph{total} stellar mass, \Mstel. 
Nevertheless, uncertainties in our estimates of \Mbh\ could affect our measured Eddington ratios and may bias our measurement of $\gamma_E$ or indeed our conclusion that the prevalence of AGNs is independent of the stellar mass of the host galaxy and primarily determined by the Eddington ratio distribution. 
It is perhaps surprising given these large uncertainties and systematic biases in our estimates of Eddington ratios that we do recover a universal Eddington ratio distribution with a constant slope that accurately describes our data throughout our redshift range and over several orders of magnitude in \Mstel\ and \Lamedd.

However, we can disregard any uncertainties in our inferred black hole masses if we consider our \Lamedd\ values as a tracer of the \emph{specific accretion rate}---the rate of black hole growth relative to the \emph{stellar mass} of the host galaxy---rather than the true Eddington ratio.
The results shown in Figure \ref{fig:leddfrac_massbin} show that the probability of finding an AGN with a \emph{specific accretion rate} is described by a power-law distribution that is independent of stellar mass. 
Thus AGNs are \emph{not} predominantly found in massive galaxies when we consider AGNs growing above a \emph{specific accretion rate}.
The conclusions laid out above and in Section \ref{sec:discuss} below are therefore robust despite the significant uncertainties in our estimates of \Mbh.

We note that evolution of the $\mbh-\sigma$ relation could alter the slope of the redshift evolution, $\gamma_z$, that we find for $p(\lamedd \giv \mstel,z)$. 
Indeed, \citet{Peng06b} find that black hole masses at $z\sim1$ are a factor $\sim2$ larger than implied by the local $\mbh-\sigma$ relation. 
However, significantly stronger evolution of the $\mbh-\sigma$ relation is required to account for the entire factor $\gtrsim 10$ evolution we find in $p(\lamedd \giv \mstel,z)$ between $z\sim0.2$ and $z\sim1$. 

In the remainder of this paper we discuss results in terms of Eddington ratio, but note that we actually measure the \emph{specific accretion rate} (with a particular scaling factor) and thus our results could instead be interpreted in terms of this quantity, providing more freedom in the underlying distribution of black hole masses.

\begin{table}
\caption{Best-fit Parameters from Maximum-likelihood Fitting with Redshift Evolution and Eddington Ratio Dependence}
\begin{center}
\begin{tabular}{c c}
\vspace{-5mm}\cr
\hline
\hline
\vspace{-2 mm} \cr
Parameter & Value\\
\vspace{-2 mm} \cr
\hline 
\vspace{-2 mm} \cr
$\log_{10} A$ & $-$\amend{3.15}$^{+0.08}_{-0.08}$ \\
$\gamma_E$  & $-$\amend{0.65}$^{+ 0.04}_{-0.04}$ \\
$\gamma_z$   & \phs\amend{3.47}$^{+0.49}_{-0.48}$ \\
    \vspace{-2 mm} \cr    
\hline
\end{tabular}
\end{center}
\label{tab:eddmlfit}
\end{table}


\section{The dependence of AGN activity on host galaxy color}
\label{sec:col}

As discussed in the introduction, AGN accretion activity and star formation processes appear to be connected.
Both the overall star formation rate density and the AGN luminosity density have declined rapidly since $z\sim1$, indicating that these processes may be triggered by a common mechanism.
The strong correlation between black hole mass and the masses of central stellar bulges \citep[traced by the $\mbh-\sigma$ relation;][]{Ferrarese00,Gebhardt00,Tremaine02} also implies that these processes are intertwined. 
Whether AGNs play a direct role in the evolution of galaxies, for example, regulating star formation via feedback processes, is a vital, unresolved question.

In this section we investigate the positions of AGNs in the optical color--magnitude diagram and determine the prevalence and distribution of AGN activity within the two prominent galaxy populations---blue cloud galaxies, associated with ongoing star formation activity; and the red sequence of generally older, quiescent, and more massive galaxies---as well as the green valley that lies in between and may represent a transition population
\citep[e.g.,][]{Martin07,Salim09,Mendez11}.
Indeed, an overdensity of AGNs in the luminous galaxies close to and within the green valley has been observed \citep{Nandra07b,Coil09,Hickox09} and may be evidence that AGN activity is associated with the color transformation of galaxies. 
Recent work by \citet{Xue10}, however, has shown the importance of stellar-mass selection effects in such analyses, indicating that stellar mass, not color, may be the key parameter that drives the observed trends in the color-magnitude diagram. 
In Sections \ref{sec:frac} and \ref{sec:zevol} above we show that $p(\lx \giv \mstel,z)$ is strongly dependent on stellar mass, redshift, and X-ray luminosity 
(although the underlying trend is likely driven by Eddington ratio, see Section \ref{sec:eddratio}). 
These trends will significantly affect the observed distribution of AGNs in the color--magnitude diagram. 
Thus, we must account for these effects to investigate the true connection between the prevalence of AGN activity and the colors of potential host galaxies, allowing us to shed light on the relationship between SMBH accretion and star formation processes.

\subsection{AGNs in the Color--magnitude Diagram}
\label{sec:cmd}

\begin{figure}
\begin{center}
\includegraphics[width=0.45\textwidth]{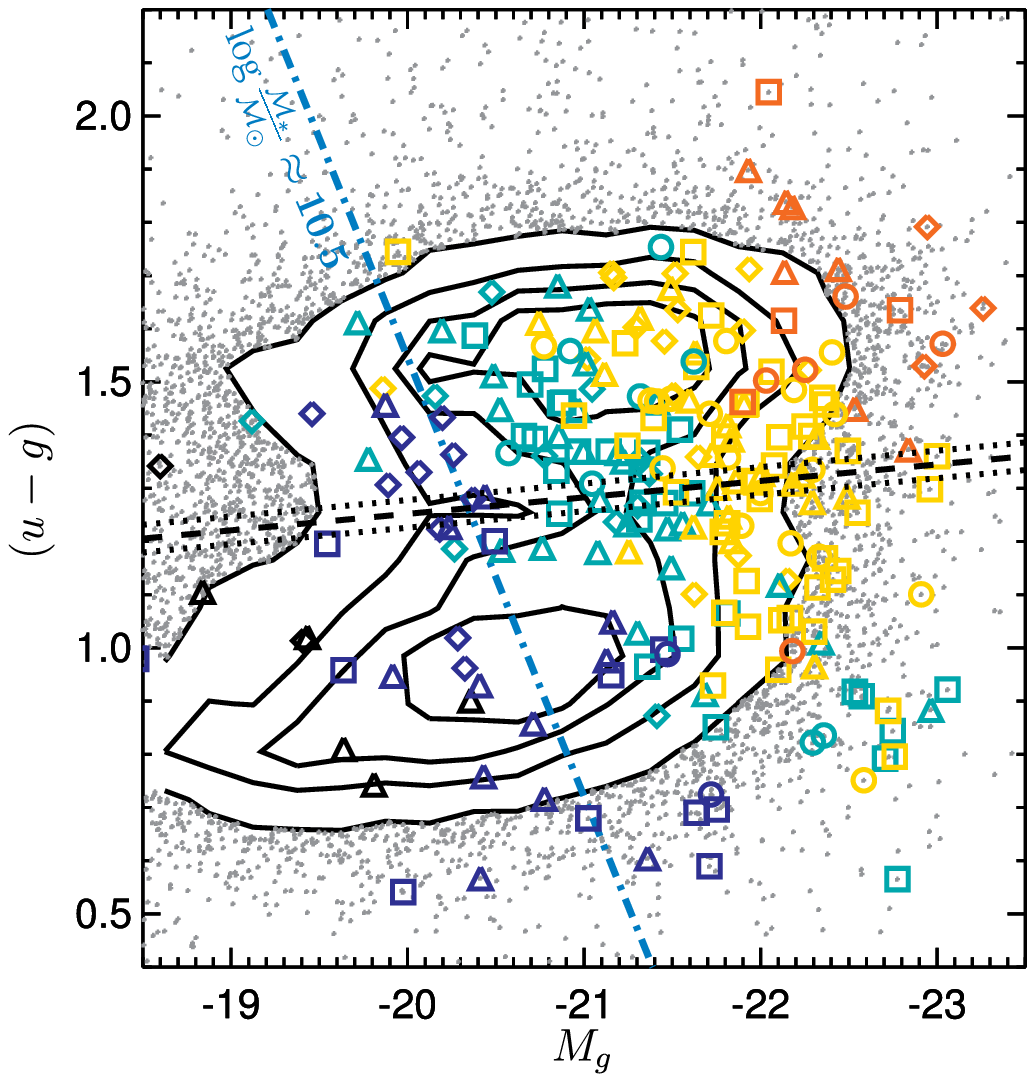}
\vspace{0.1cm}\\
\includegraphics[width=0.45\textwidth]{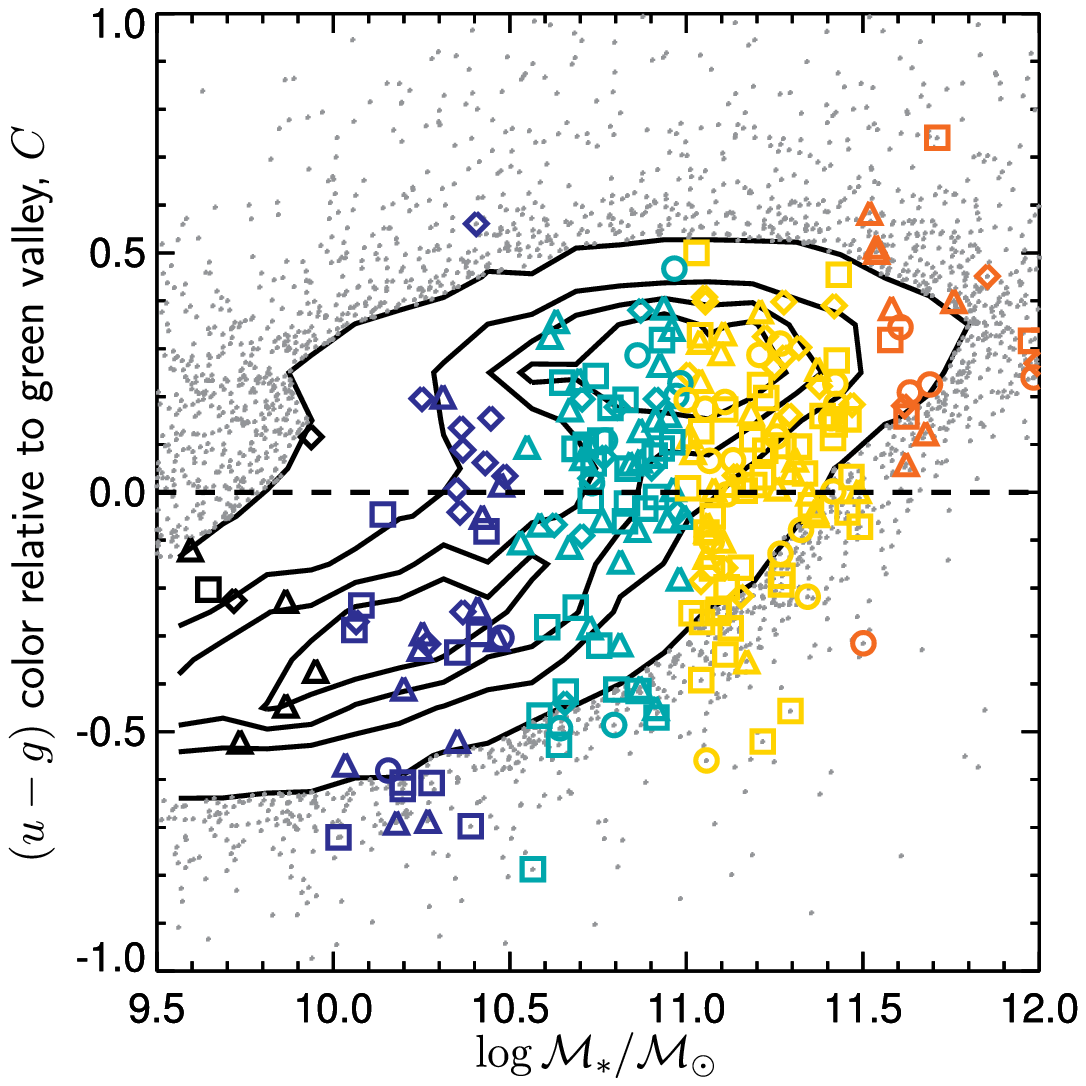}
\end{center}
\caption{
\textit{Top:} color--magnitude diagram for our parent sample of galaxies (small gray dots and contours) and X-ray-detected AGN sample (colors and symbols correspond to the mass and luminosity bins used in Figures \ref{fig:fraclxmass} and \ref{fig:bestfrac}). 
The green valley evolves over the redshift spanned by our sample: the dashed line shows the center of the green valley at the center of our redshift range, $z=0.6$ (from Equation \ref{eq:coldiv}); the dotted lines show the green valley at $z=0.2$ and $z=1.0$. 
 The light blue dot-dashed line shows the track in color--magnitude space based on the simple linear relations  between mass-to-light ratios and $(u-g)$ color given by \citet{Bell03} for $\log \mstel/\mathcal{M}_\odot=10.5$. 
Our mass estimates, which use multiple photometric bands and redshift information, do not directly follow this linear relation, and thus cut through the color-magnitude diagram in a complex fashion.
\textit{Bottom:} 
as in the top panel, but converted to the rest-frame $(u-g)$ color relative to the green valley (for the magnitude and redshift of a given source) on the $y$-axis and \Mstel\ on the $x$-axis. 
}
\label{fig:cmd}
\end{figure}

Figure \ref{fig:cmd} (top) shows the color--magnitude diagram (after applying the redshift-dependent stellar-mass cut described in Section \ref{sec:stellarmasslim} and shown in Figure \ref{fig:mass_vs_z}) for our parent sample of PRIMUS galaxies and our X-ray-detected AGN samples. 
The color distributions of all galaxies change over the redshift range spanned by our sample; at higher redshift the position of the red sequence and blue cloud shift to bluer colors \citep[e.g.,][]{Bell04}. 
Additionally, the green valley that divides these populations has a weak magnitude dependence, following the slope of the red sequence (as can be seen in Figure \ref{fig:cmd}, top).
Stellar mass limits also cut diagonally through the color--magnitude diagram \citep{Bell03}, as demonstrated by the blue dashed line in Figure \ref{fig:cmd} (top).
To cleanly separate blue and red galaxies and reveal any possible overdensity of AGNs within the intermediate population, we choose to present results based on the color relative to the green valley, $C$, where we define the green valley as a function of $M_g$ and redshift using Equation \ref{eq:coldiv}. Thus,
\begin{equation}
C = (u-g) - (0.671 - 0.031 M_g - 0.065 z)
\end{equation}
where $(u-g)$ is the rest-frame color of a galaxy and $M_g$ and $z$ are the absolute $g$-band magnitude and redshift.
We re-project the color--magnitude diagram to the more informative $C$ versus \Mstel, as shown in the bottom panel of Figure \ref{fig:cmd}.
It is immediately apparent that our \emph{detected} AGN sample is predominantly in the more massive galaxies, consistent with the increase 
in $p(\lx \giv \mstel,z)$ found in the preceding sections. 
They do, however, appear to span the full range of galaxy colors at any given stellar mass.

\subsection{The fraction of AGN as a function of host galaxy color}
\label{sec:frac_vs_col}

\begin{figure*}
\includegraphics[width=\textwidth]{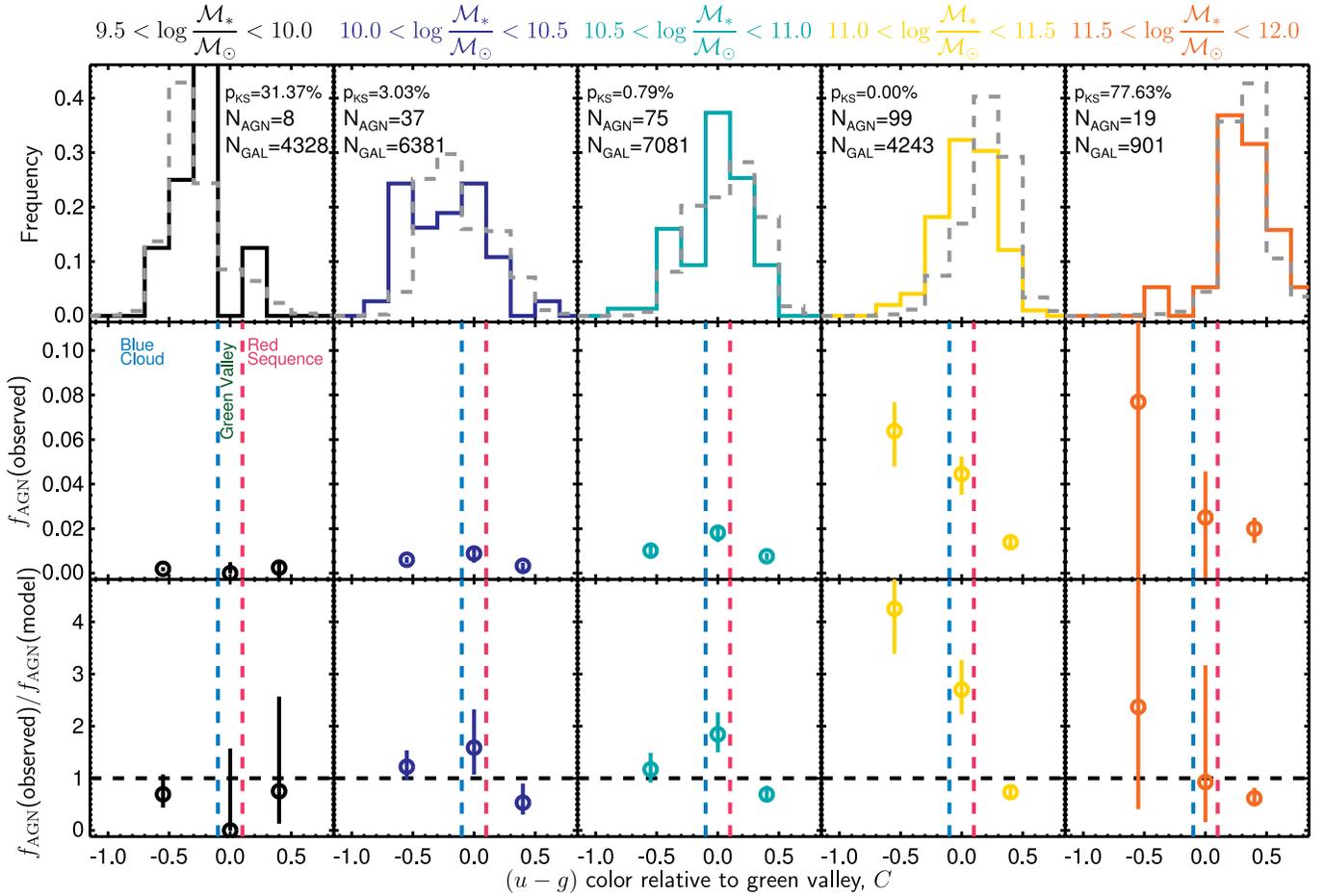}
\caption{
\textit{Top panels}: histograms of the color (relative to the green valley) of the X-ray-detected AGN samples (colored lines) and parent galaxy sample (gray dashed lines) in our five stellar mass bins. The histograms are normalized to show the fraction of sources within a given population in each color bin.
In each panel we give the total number of galaxies and X-ray-selected AGNs in that stellar mass bin; we also show the probability, based on the Kolmogorov--Smirnov test, that the two samples are drawn from the same distribution. 
In the highest two \Mstel\ bins we find significant evidence that the underlying distributions are different---AGNs appear to be associated with bluer host galaxies. For the lower \Mstel\ bins we cannot exclude the possibility that the AGNs are drawn from the same distribution of colors as the parent galaxy sample.
\textit{Middle panels:} the observed fraction of parent galaxies found to host X-ray AGNs (i.e., the ratio of source numbers used in the histograms in the top panels) in three color bins corresponding to the blue cloud, green valley, and red sequence. No correction is performed to account for the X-ray incompleteness, redshift evolution, or dependence of the AGN fraction on \Lx\ or \Mstel.
\textit{Bottom panels:} the ratio of the observed fraction of galaxies hosting AGNs to that expected based on our model for $p(\lx \giv \mstel,z)$ and accounting for our X-ray incompleteness. We find that the observed fraction of AGNs is higher than expected from our model in the green valley and blue cloud, and lower than expected in the red sequence.
}
\label{fig:colfrac}
\end{figure*}

In the top panels of Figure \ref{fig:colfrac} we show histograms of $C$ for our parent galaxy samples (dashed gray lines) in five stellar mass bins, compared to the histograms for the X-ray-detected AGN sub-sample (solid colored lines). 
In two of our stellar mass bins we find that the color distributions of the AGN host galaxies and the overall galaxy population are significantly different based on a Kolmogorov--Smirnov (KS) test, which gives probabilities of 0.79\% and 0.00\% that the AGN and galaxy samples are drawn from the same color distributions for the $10.5<\log \mstel/\msun<11.0$ and $11.0<\log \mstel/\msun<11.5$ stellar mass bins, respectively.
The colors of the AGN host galaxies are in general bluer than the overall galaxy population.
In the other stellar mass bins the K-S test finds that there is no significant evidence (at the 99\% confidence level) for any difference in the color distributions, although this may be partly due to the smaller samples of AGNs.

The middle panels of Figure \ref{fig:colfrac} show the observed fraction of galaxies with an X-ray-detected AGN in three color bins corresponding to the blue cloud, green valley and red sequence. 
The strong increase in the observed fraction of AGNs as \Mstel\ increases dominates the behavior resulting in an overall increase in the AGN fraction for all colors at higher stellar masses, but we also see evidence for an increased AGN fraction in bluer host galaxies.

The bottom panels show the observed fraction of galaxies hosting an AGN relative to the fraction predicted by our model (with no color dependence) 
for $p(\lx \giv \mstel,z)$ determined in Section \ref{sec:zevol}.
This accounts for the strong increase in $p(\lx \giv \mstel,z)$ with stellar mass and redshift and corrects for the effects of the X-ray sensitivity, 
allowing us to compare the model with the observed data in different color bins and identify deviations that depend on color.
In our lowest stellar mass bin we find no evidence for a deviation from the best-fit relation, but in the remaining bins there is evidence for a weak enhancement of AGN activity in the blue cloud and green valley compared to the red sequence, by a factor $\sim 2-4$.
There are hints that the enhancement is greater in the green valley compared to the blue cloud for $10.0<\log \mstel/\msun<11.0$ (blue and green points), whereas at $11.0<\log \mstel/\msun<11.5$ (yellow points) the enhancement is greater in the blue cloud. 
Given the uncertainties in individual measurements, however, there is not significant evidence of a difference between the enhancement factor for blue cloud and green valley galaxies, or for a joint \Mstel\ and color dependence.
In our lowest and highest stellar mass bins there are too few objects to identify any color dependence. 
\moreamend{We also note that due to our redshift-dependent stellar mass limits (see Wection \ref{sec:stellarmasslim}) our lower stellar mass samples ($\log \mstel/\msun<10.5$)
are restricted to the lower end of our redshift range ($z\lesssim 0.6$).
Redshift evolution of the overall AGN fraction (as determined in Section \ref{sec:zevol}) is accounted for in the lower panels of Figure
\ref{fig:colfrac}, but the results could be biased if the relative enhancement in the blue cloud and green valley evolves differently to the total population.
}

\subsection{Maximum-likelihood Fitting in Multiple Color Bins}
\label{sec:ml_col}

\begin{figure}
\begin{center}
\includegraphics[width=0.4\textwidth]{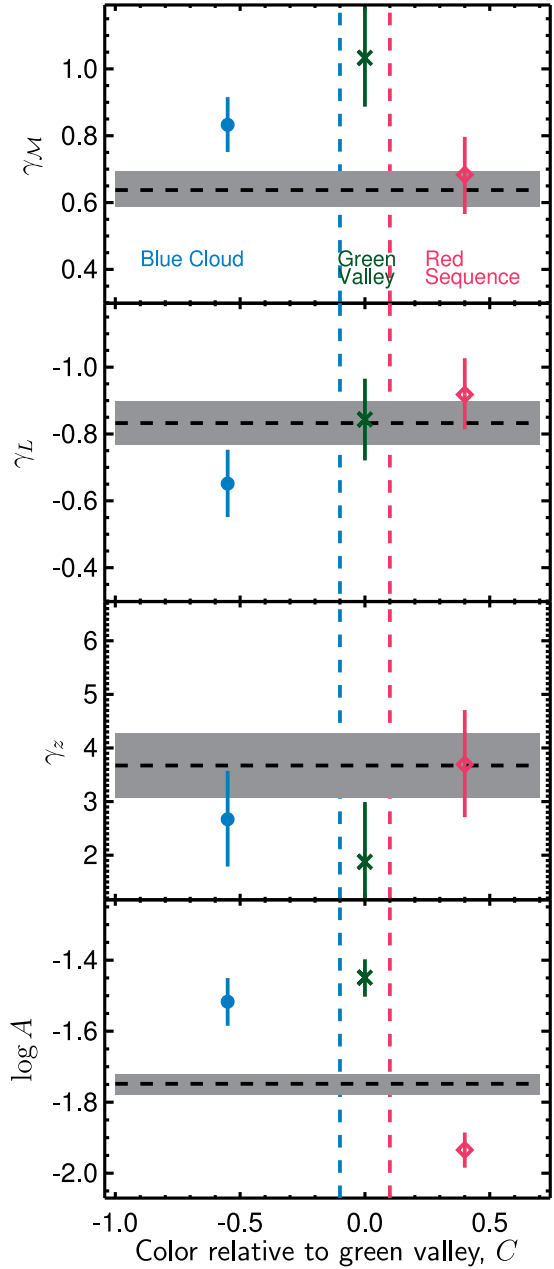}
\end{center}
\caption{
Best-fit parameters from the maximum-likelihood fitting for $p(\lx \giv \mstel,z)$ in three color bins corresponding
to the blue cloud, green valley, and red sequence, compared to overall best fit parameters and their uncertainties
from Section \ref{sec:zevol} (horizontal black dashed lines and gray regions).
There is weak evidence that the parameters describing the power-law dependences on \Lx, \Mstel and $z$ 
depend on the galaxy color. 
We find very strong evidence that the overall normalization is higher for the blue cloud and green valley than red sequence by a factor of $\sim 2$.
The prevalence of AGN activity appears to be mildly enhanced in galaxies with increased star formation activity.
}
\label{fig:parsvscol}
\end{figure}

To quantify these trends, we have repeated our maximum-likelihood fitting (Section \ref{sec:mlfit}) to constrain the functional form of $p(\lx \giv \mstel,z,C)$, which describes the probability an AGN of luminosity \Lx\ being found in a galaxy with a given stellar mass, redshift and color.
We assume the same power-law dependence on \Lx, \Mstel\ and $z$ given by Equation \ref{eq:zevol}.
Figure \ref{fig:parsvscol} shows the best-fit parameters in the three bins of galaxy color corresponding to the blue cloud, green valley and red sequence, compared to the overall best fit from Section \ref{sec:inc_zevol}.
There is weak evidence that the three power-law slopes ($\gamma_\mathcal{M}$, $\gamma_L$ and $\gamma_z$) are dependent on galaxy color. 
The slope controlling the stellar mass dependence ($\gamma_\mathcal{M}$) may be steeper in the blue cloud, and the luminosity dependence may have a flatter slope ($\gamma_L$), indicating a higher prevalence of AGNs in blue galaxies with high stellar masses and a tendency for blue galaxies to host more luminous AGNs.
Nevertheless, all the slopes are consistent with our overall best-fit relation to within $<3\sigma$.
Thus, while there are hints of a more complex behavior, the shape of the overall
trends in the prevalence of AGN activity as a function of stellar mass, X-ray luminosity and redshift can be well described by the functional form derived in Section \ref{sec:zevol}.
The bottom panel of Figure \ref{fig:parsvscol}, however, shows that our best-fit normalizations are significantly higher for the blue cloud and green valley than  for the red sequence. 
This is consistent with the behavior found in Section \ref{sec:frac_vs_col} above and seen in Figure \ref{fig:colfrac}.
The stellar mass dependence drives the trends seen in the color--magnitude diagram, and thus more AGN are found in massive galaxies, which tend to have redder colors, but once this effect is accounted for we have shown that AGN activity is enhanced in galaxies with \emph{bluer} colors associated with increased star formation activity.

\subsection{AGN contamination of the host galaxy light}
\label{sec:agncontamination}

We note that the optical light of some of our X-ray AGN host galaxies may be contaminated by the AGN, which could make the observed SED appear bluer than that of the actual galaxy.
In these cases our derived stellar masses should still be robust as the erroneous addition of a young stellar component to account for the blue light will contribute relatively little additional stellar mass.
However, AGN contamination could skew the observed color distribution and contribute to the observed enhancement of AGN activity in the blue cloud and green valley.
In the most extreme case, an unobscured AGN that contributes a fraction $\gtrsim 30$\% of the $B$-band optical light in a red, elliptical host galaxy would skew the observed colors by up to $\Delta(u-g)\approx1.0$, easily moving the galaxy from the red sequence into the green valley or blue cloud \citep{Pierce10b}.
We have identified and excluded sources that exhibit broad emission lines in their optical (PRIMUS) spectra, 
which will exclude the most severely contaminated objects where an unobscured AGN is clearly dominating over the host galaxy optical light. 
We also set an upper limit on the X-ray luminosity of $\lx=10^{44}$ \ergs, which prevents the inclusion of luminous AGNs that are also likely to have biased observed colors.
Nevertheless, AGN contamination may still affect the observed colors of our sample.
Recent work by \citet{Pierce10} comparing the nuclear and outer colors of X-ray-selected AGN hosts at $z\sim1$ showed that in general the integrated colors were unaffected, except in rare cases ($<10$\% of galaxies) that are X-ray bright ($\lx>2\times 10^{43}$ \ergs) and appear unabsorbed at X-ray wavelengths. 
Thus, AGN contamination is not expected to severely bias our overall results, but could affect our highest luminosity bins.
To check the extent of any bias we artificially redden the colors of all X-ray-detected AGNs in our highest luminosity bin ($\lx>10^{43.5}$ \ergs) and in the blue cloud ($C<-0.1$) by $\Delta(u-g)=1.0$, and repeat our maximum-likelihood fitting.
We find that the $\gamma_L$, $\gamma_\mathcal{M}$, and $\gamma_z$ slopes are altered but remain within their statistical error bars.
The overall factor $\sim2$ enhancement in the green valley and blue cloud is not affected.
Thus, any systematic bias will not affect our conclusions even in this extreme case of severe AGN contamination for all of our most luminous sources.

\subsection{Eddington ratio distribution for different host galaxy colors}

\begin{figure}
\begin{center}
\includegraphics[width=0.4\textwidth]{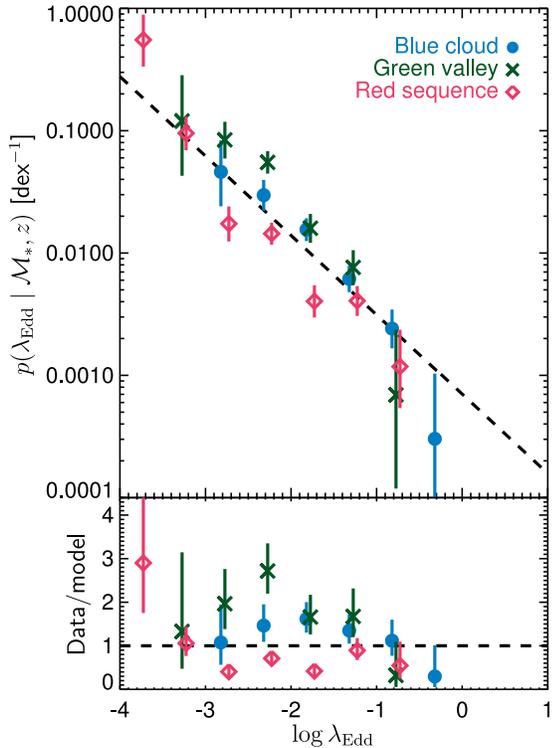}
\end{center}
\caption{
Distribution of Eddington ratios, $p(\lamedd \giv \mstel,z,C)$, for galaxies in the blue cloud, green valley, and red sequence (top). 
The bottom panel shows the ratio of the data in the top panel to our best-fit model from Section \ref{sec:eddratio}. 
We find a slightly higher fraction of galaxies hosting AGNs in the blue cloud and green valley (factor $\sim2$), although the distribution of Eddington ratios has approximately the same shape, with AGNs found over the full range of Eddington ratios in galaxies of all colors.
}
\label{fig:leddfrac_colbin}
\end{figure}

In Figure \ref{fig:leddfrac_colbin}, we show the distribution of Eddington ratios for our three color bins, again using the $N_\mathrm{obs}/N_\mathrm{mdl}$ method to compare the observed fraction with that predicted by our best-fit relation determined in Section \ref{sec:eddratio}. 
The results also show an enhancement of $p(\lamedd \giv \mstel,z)$ in the green valley and blue cloud compared to the red sequence (significant at $>3 \sigma$ when combining over the full \Lamedd\ range). 
However, the shape of the distribution of Eddington ratios is roughly the same for all colors, with approximately the same slope and AGNs found throughout the full observable range of $\lambda_\mathrm{Edd}\approx10^{-4}-10^{-1}$. 
AGN activity is clearly enhanced in galaxies that tend to have bluer colors (indicating active star formation is taking place), but the uniform distribution of Eddington ratios is strong evidence that the basic physical processes that trigger and fuel the AGN are essentially the same. 
It should be noted that the color-dependent enhancement is relatively weak (factor $\sim 2$) compared to the overall trend with Eddington ratio \emph{or} the strong redshift dependence.
We discuss our findings and their implications further in Sections \ref{sec:hostcolor} and \ref{sec:role} below.

\section{Discussion}
\label{sec:discuss}

The results presented in this paper reveal how the prevalence and distribution of  
moderately obscured, moderate-luminosity AGN accretion activity 
is related to the properties of their potential host galaxies during a key epoch in the lifetime of the universe, $0.2<z<1.0$.
We find that the distribution of X-ray luminosities for all stellar masses has the same power-law shape. 
The probability of a galaxy hosting an AGN above a given \emph{X-ray luminosity} increases strongly with the stellar mass of the galaxy. 
However, we show that this effect may be primarily driven by the Eddington ratio distribution.
The distribution of Eddington ratios has a power-law shape; a low fraction of galaxies ($\sim 0.1$\%) host AGNs accreting close to the Eddington limit (\Lamedd$\gtrsim0.1$), whereas $\sim 10$\% of galaxies contain an SMBH growing at a very low rate (\Lamedd$\sim 10^{-4}$). 
The probability of hosting an AGN of a given Eddington ratio is independent of stellar mass. 

Alternatively, we can interpret these results in terms of a specific accretion rate: the rate of black hole growth relative to the stellar mass of the galaxy.
The distribution of specific accretion rates is a universal function that is independent of stellar mass (see Section \ref{sec:specaccrate}).
Our conclusions hold regardless of uncertainties in the black hole mass and thus the absolute value of the Eddington ratios for the AGN.

We find that AGNs are \emph{not} predominantly in massive, red galaxies.
More AGNs may be found in red galaxies due to a combination of two selection effects:
(1) it is easier to detect the more prevalent, low Eddington ratio (low specific accretion rate) AGNs in galaxies with higher stellar masses as the AGNs will be intrinsically brighter;
and (2) more massive galaxies tend to have redder colors.

There is, however, a mild dependence of the prevalence of AGN activity on galaxy color.
While AGNs are found in galaxies across the whole range of colors, there appears to be an 
enhancement (factor $\sim 2$) in the probability of hosting an AGN for galaxies with \emph{blue} colors, 
both within the blue cloud and the green valley.
Thus there may be a weak association between AGN activity and star formation within a host galaxy. 
Nevertheless, a lack of star formation does not preclude AGN activity---the distribution of Eddington ratios in red galaxies follows approximately the same power-law shape as in blue galaxies, spanning the full range of accretion rates.

AGN activity also evolves strongly with redshift.
We find that the probability of a galaxy hosting an AGN has decreased rapidly since $z\sim1$,
but the distribution of Eddington ratios retains the same power-law shape. 
Thus, the same physical process or processes are 
likely to be responsible for triggering and fueling AGNs throughout this epoch, 
but they must diminish rapidly over the time period
 (see Section \ref{sec:agntriggering}).

In this section we discuss the implications of our results. 
In Sections \ref{sec:eddratiohighz} and \ref{sec:eddratiolocal} we compare our findings with previous studies of the Eddington ratio distribution at similar redshifts to our study and in the local universe respectively. 
In Section \ref{sec:agntriggering} we discuss the implications of our results for models of AGN triggering and fueling processes over the last 8 billion years. 
In Section \ref{sec:hostcolor} we discuss our findings on the dependence of AGN accretion activity on host-galaxy color and compare with prior studies of color--magnitude relations for AGNs. 
Section \ref{sec:role} discusses the role of AGNs in galaxy evolution.
Section \ref{sec:issues} discusses the remaining issues and uncertainties in our work and prospects for future studies.

\subsection{Comparison with prior studies of the Eddington ratio distribution at $z\gtrsim0.2$}
\label{sec:eddratiohighz}

Several prior studies have presented distributions of Eddington ratios for X-ray-selected samples of AGNs at $z\gtrsim0.2$.
\citet{Hickox09} found that X-ray-selected AGNs in AGES at $0.25<z<0.8$ have a wide range of Eddington ratios spanning $10^{-3}<\lamedd<1$, with a broad, roughly lognormal distribution peaking at $\lamedd\approx0.02$. However, the incompleteness effects due to the X-ray sensitivity were not taken into account. 
\citet{Simmons11} determined Eddington ratios for X-ray sources in the GOODS fields (using very deep X-ray data from both the \chandra\ Deep Fields) at $0.25<z<1.25$ and found that the sample was dominated by objects accreting at very low Eddington ratios ($\lamedd<0.01$) with a median of $\lamedd=0.006$; the extremely deep X-ray data mitigate incompleteness effects and enable detection of very low Eddington ratio sources.
\citet{Brusa09b} also presented an Eddington ratio distribution for obscured, moderate-luminosity, X-ray AGNs identified in the \chandra\ Deep Field-South at higher redshifts: $1<z<2$ and $2<z<4$. Similar, broad distributions of Eddington ratios were found in both redshift bins with medians $\lamedd\sim0.02-0.08$.
None of these prior papers have tracked or corrected for X-ray selection effects to reveal the underlying distribution of Eddington ratios.

Recently, \citet{Trump11} proposed that there are two distinct distributions of Eddington ratios for X-ray-selected AGNs corresponding to two different modes of accretion. 
They find that broad-line QSOs are predominantly at higher \Lamedd$\sim0.1$, consistent with other studies of the Eddington ratio distribution of QSOs \citep[e.g.,][]{Kollmeier06,Gavignaud08,Kelly10}.
They propose that these AGNs are fed by a thin accretion disk with a disk wind containing the broad-line region and some obscuring material. 
Narrow-line AGNs or X-ray AGNs lacking optical AGN signatures are shown to have lower \Lamedd$\sim10^{-3}$ and may be powered by a geometrically thick, radiatively inefficient accretion flow.
Our sample was explicitly limited to ``obscured" AGNs that lack broad lines in their optical spectra (allowing us to probe the host galaxy properties in the optical light) and will thus correspond to the second class proposed by \citet{Trump11}. 
Consistent with their findings, we have shown that these AGNs are predominantly at low Eddington ratios.
In fact, we find a power-law relation that extends to very low \Lamedd$\sim10^{-4}$.
Therefore, the physical processes that trigger and fuel low-level AGN accretion activity \amend{may} also be common and take place in galaxies of all stellar masses.

Our results are consistent with previous studies of Eddington ratio distributions for moderately obscured, moderate-luminosity AGNs at $z\gtrsim0.2$.
However, we show that the true distribution of Eddington ratios is a power law that continues to rise to lower values of \Lamedd. 
This result is reliant on careful analysis of X-ray sensitivity effects relative to the stellar masses of the host galaxies.
Our directly observed distribution of Eddington ratios has a broad, approximately lognormal distribution with a peak at \Lamedd$\sim10^{-2}$ (see gray histogram in final panel of Figure \ref{fig:leddfrac_massbin}) as found in prior works but this is significantly skewed by X-ray incompleteness effects and does not represent the true \Lamedd\ distribution.
We note that our study has excluded BLAGNs which may dominate the population at high \Lamedd\ and could correspond to a distinct accretion mode.

\subsection{Comparison with prior studies of the local Eddington ratio distribution}
\label{sec:eddratiolocal}

The lack of an extremely large area, sensitive X-ray survey at $>2$ keV has meant that the majority of studies of the Eddington ratios of AGNs in the local universe ($z\lesssim0.1$) have relied on large-area optical spectroscopic surveys to identify AGNs from their optical line-emission, in particular the Sloan Digital Sky Survey \citep[SDSS; e.g.,][although see also 
\citealt{Georgakakis11}]{Kauffmann03,Heckman04,Kauffmann04,Hao05,Kewley06,Schawinski07}. 
As discussed in Section \ref{sec:intro}, such studies have found that the fraction of galaxies hosting optically selected narrow-emission-line (Type 2) AGNs rises steeply with increasing stellar mass and that Type 2 AGNs are found predominantly in massive galaxies  
\citep[$\mstel\gtrsim10^{10} \msun$;][]{Kauffmann03,Best05}.

\citet[hereafter H04]{Heckman04} studied the present-day growth rates of black holes in SDSS galaxies using the O{\sc iii} luminosity as a tracer of AGN accretion activity. 
They found that the distribution of Eddington ratios is highly dependent on black hole mass; while $\sim0.5$\% of black holes with $\mbh=10^{7}\msun$ are accreting at Eddington ratios $\lamedd>0.1$, less than 0.01\% of higher mass black holes ($\mbh\approx3\times10^{8}\msun$) are accreting at these high Eddington ratios. 
\citet[hereafter K09]{Kauffmann09} built on this work and proposed that black hole growth in the local universe is driven by two distinct processes:
(1) a self-regulated mode of growth in young galaxies with plentiful cold gas that leads to a broad lognormal distribution of Eddington ratios peaking at \Lamedd$\sim10^{-2.5}$,
and (2) a passive mode in galaxies with older central stellar populations where the supply of fuel to the black hole is driven by stellar mass loss from  stars, resulting in a power-law distribution of Eddington ratios.
This is somewhat different to our results at higher redshifts using X-ray selection but probing similar ranges of Eddington ratios and black hole masses (roughly traced by the galaxy stellar mass).
We find that the distribution of Eddington ratios is the same for all stellar masses and is described by a power law; we find no evidence for an additional lognormal component and that the shape of the Eddington ratio distribution is similar for young and old stellar populations (i.e., blue and red galaxies).

It is unclear how to reconcile our findings with the results of H04 and K09 without further study to directly compare X-ray and optical probes of AGN accretion in both the local and high-redshift universe. 
The discrepancies could be related to biases inherent to the two very different selection techniques that affect both the completeness of the AGN samples and the estimates of the Eddington ratios. 
For example, AGNs in the lognormal mode identified by K09 could be heavily obscured and thus are not detected in X-rays or are detected but their luminosities (and thus Eddington ratios) are significantly underestimated.
Alternatively, it is possible that the O{\sc iii} emission line may not provide an accurate tracer of the black hole accretion at low Eddington ratios or in galaxies with current star formation. 

Another possibility is that the differences between our results and those of H04 and K09 are due to evolution with redshift.
Indeed, H04 consider volume-averaged growth rates and show that the most massive black holes in their sample are \emph{not} growing at significant rates at the present day and \emph{must} have undergone periods of rapid growth in the earlier universe. 
At higher redshifts we may be seeing this earlier phase of relatively rapid growth for all galaxies over the range of stellar masses (and thus black hole masses) that we are able to probe, although our results do not show any direct evidence for such ``downsizing'' behavior (see further discussion in Section \ref{sec:agntriggering} below).

The relative importance of the two accretion modes proposed by K09 could also change with redshift.
We find that the distribution of Eddington ratios has a power-law shape with a normalization that increases rapidly with redshift as $\sim(1+z)^{3.5}$.
The lognormal accretion mode seen by K09 may not evolve or may evolve only weakly with redshift and thus would not be seen in our sample. 
This may, however, conflict with the K09 interpretation that the accretion is driven by stochastic accretion of cold gas, as the supply of cold gas should be more plentiful in the high-redshift universe. 

Further work and a direct comparison of X-ray and optical probes of AGN accretion modes in both the local and high-redshift universe is required to resolve these issues.

\subsection{The Evolution of AGN Triggering and Fueling}
\label{sec:agntriggering}

Our observed AGN Eddington ratio distribution can provide key insights into the physical processes that are responsible for triggering and fueling AGNs at $z\sim0.2-1$.
One of our major findings is that the probability of hosting an AGN for galaxies throughout our stellar mass range ($9.5<\log\mstel/\msun<12$) is primarily determined by the Eddington ratio distribution and is independent of the stellar mass.
Not only is the shape of the distribution of accretion rates (tracked by the Eddington ratios) the same, the overall normalization does not depend on the stellar mass.
Thus, AGN activity may take place in any galaxy, regardless of the stellar mass. 
The universality in both the prevalence of black hole growth and the distribution of accretion rates (Eddington ratios) indicates that the same underlying physical processes are responsible for triggering and fueling AGN activity in all moderately massive galaxies. 

We also find that the fraction of galaxies hosting AGNs above a given Eddington ratio evolves strongly with redshift with the form $(1+z)^{3.5\pm0.5}$ to $z=1$. 
The shape of the distribution remains the same, indicating that the same processes are regulating AGN activity but are undergoing a rapid evolution with redshift.  
This evolution could take two possible forms:  
(1) the fraction of accreting systems is higher at higher redshift, 
or (2) the characteristic Eddington ratio is higher at higher redshift.  
In both scenarios, a given physical process must trigger AGN activity across our redshift range, resulting in the same distribution of Eddington ratios with time.  
However, in the first scenario the fraction of galaxies with this process is higher at earlier times, while in the second scenario the AGNs are more rapidly accreting at earlier times.  We cannot distinguish between these two scenarios solely with the results presented here.

However, studies of the evolution of the X-ray luminosity function (XLF) of AGNs, which traces the overall distribution of AGN accretion activity, find that at low redshifts ($z\lesssim1$) the evolution is dominated by luminosity evolution \citep{Barger05,Aird10}. 
The characteristic luminosity, $L_*$, corresponding to the break in the double-power-law form of the XLF, shifts to lower luminosities between $z\sim1$ and the present day. The Eddington ratio distribution presented here will track the evolution of the faint end of the XLF. 
Indeed, if we translate the luminosity evolution of $L_*$ into an effective change in normalization at the faint end of the XLF we find evolution of the form $(1+z)^{3.9\pm0.3}$ \citep[based on the ][ ``luminosity and density evolution" model]{Aird10}, in excellent agreement with our evolution in the normalization of the Eddington ratio distribution presented above. 
We therefore conclude that our redshift evolution is the result of a shift in the accretion processes to lower characteristic Eddington ratios since $z\sim1$. 

This implies that any upper boundary to the Eddington ratio distribution would also move to lower \Lamedd\ as redshift decreases. 
Our results do not show evidence for such a redshift-dependent cutoff, although we probe only relatively low Eddington ratios, especially at low redshifts (see Figure \ref{fig:leddfrac_zbin}), and do not include broad-line QSOs that may dominate the population at the highest Eddington ratios.
Much larger samples are required to determine the presence and evolution of an upper boundary or turnover of our Eddington ratio distribution at high \Lamedd.
In fact, \citet{Steinhardt10} have presented evidence for a ``sub-Eddington boundary" in large samples of SDSS quasars at $0.2<z<4$, which depends on black hole mass and may evolve with redshift, although the connection to our results requires further investigation.

Our findings do not agree with the ``downsizing" picture of AGN evolution, which attributes the luminosity evolution of the XLF to a decrease in the average mass of galaxies that host AGNs as time progresses, either due to a shut off of AGN activity in the highest mass galaxies or AGNs in lower mass galaxies ``turning on" at later times \citep[e.g.,][]{Barger05,Alonso-Herrero08,Hopkins09b}.
This picture is not supported by our results:
we find that the distribution of accretion rates is uniform for galaxies of all stellar masses at any given redshift.
 The entire distribution evolves with redshift, which  
we interpret as a shift to lower characteristic accretion rates with increasing cosmic time.
Thus, at any given epoch, black hole growth is taking place in galaxies over the full range of stellar masses, but at later times the accretion rates are simply much lower, on average, and there is a relative lack of the most rapidly accreting sources.
The evolution of the XLF is driven by this shift in accretion rate rather than the downsizing of black hole growth \citep[see also][]{Shankar09}.
However, as our results track the specific accretion rate relative to the host stellar mass (see Section \ref{sec:specaccrate}), we are unable to constrain any underlying evolution of the distribution of black hole masses, which may evolve differently from the stellar masses of their host galaxies and could potentially exhibit a downsizing behavior. 

While our results provide important constraints, it is currently unclear what physical mechanism could be responsible for fueling AGN activity and regulating the distribution and evolution of accretion rates.
The observed drop in the global star formation rate density from $z\sim1-2$ to $z\sim0$ \citep{Hopkins04} presumably reflects a drop in the 
availability of cold gas in galaxies to fuel star formation and potentially AGNs.
However, gas must be driven to the central region of the host galaxy to fuel an AGN.
The fueling of the AGN could be triggered by instabilities in the host galaxy caused either by secular evolution \citep[e.g.,][]{Kormendy04}, galaxy-galaxy interactions, or major or minor mergers \citep[e.g.,][]{Kauffmann00,Hopkins08}.
Alternatively AGN activity could be fed by smooth cosmological cold gas accretion \citep[e.g.,][]{Keres05}, stochastic accretion of gas within the galaxy \citep[e.g.,][]{Hopkins06c}, or mass loss from evolved stars \citep[e.g.,][]{Ciotti07,Kauffmann09}.
Whichever process controls the fueling of AGNs must be relatively independent of the stellar mass of the host galaxy, at least over the range probed by our study. 
Significant additional study is required to fully reveal the processes that are driving AGN accretion activity.

\subsection{The relationship between AGN activity and host galaxy color}
\label{sec:hostcolor}

We find that the prevalence of AGN activity is somewhat enhanced in galaxies with blue optical colors,
once stellar-mass-dependent selection effects are fully accounted for.
In addition, we show that the probability of hosting an AGN for galaxies throughout our stellar mass range is primarily determined by the Eddington ratio.
This drives the observed increase in the fraction of galaxies hosting X-ray-\emph{detected} AGNs as the stellar masses of the parent galaxy sample increase.
Thus, galaxies throughout the color-magnitude diagram may host an AGN, but a higher fraction of AGNs are \emph{detected} within the most luminous galaxies, which also tend to have redder colors.
In fact, the probability of hosting an AGN above a given Eddington ratio is higher for galaxies within the blue cloud and the green valley. 

Both \citet{Xue10} and \citet{Silverman09b} have previously investigated the fraction of AGNs as a function of host-galaxy color and stellar mass.
Indeed, 
\amend{\citet{Silverman09b} highlighted the importance of stellar-mass-dependent selection effects and the potential bias in studies using optical flux-limited samples
\citep[e.g.,][]{Nandra07b,Coil09,Hickox09,Schawinski09,Georgakakis11}, which was later demonstrated explicitly by \citet{Xue10}.
The \citet{Xue10} study also found that}
moderate-luminosity X-ray AGNs are distributed over the full range of colors in stellar-mass-matched samples with no specific clustering in the color--magnitude diagram. However, their sample of AGNs at $0<z<1$ is comparatively small (83 X-ray-detected AGNs) and was combined over this entire redshift range where we have found strong evolution of the AGN fraction. In addition, they are mainly reliant on photometric redshift estimates, which have larger errors and a higher outlier fraction than our PRIMUS redshifts, \amend{and probe to fainter optical magnitudes. 
These uncertainties will propagate to estimates of stellar masses and rest-frame colors, potentially blurring the color--magnitude diagram and masking} the relatively weak enhancement (factor $\sim2$) of the AGN fraction that we see in galaxies in the blue cloud and green valley.

\citet{Silverman09b} found that the X-ray AGN fraction at $0.5\lesssim z \lesssim1.0$ increases in galaxies with young stellar ages, traced by the $D_n(4000)$ spectral index,
and in galaxies with blue rest-frame optical colors, consistent with our results.
\citet{Silverman09b} also presented evidence that the levels of ongoing star formation, traced by the  [O{\sc ii}]$\lambda$3727 emission-line luminosity, are higher in AGN host galaxies than in non-AGN galaxies with equivalent stellar masses.

\citet[herafter S10]{Schawinski10} also studied the AGN fraction as a function of galaxy color and stellar mass.
They used a local sample of galaxies from the SDSS ($z<0.05$), identified narrow-line AGNs within these galaxies using emission-line diagnostics and determined AGN luminosities (and thus Eddington ratios) from the intensity of the [O\textsc{III}] emission line.
S10 find that the fraction of detected AGNs increases with stellar mass for all galaxies, consistent with prior studies and our findings for X-ray-detected AGNs. 
An enhancement in the fraction of AGNs is seen in the green valley along with a significant lack of emission-line AGNs in galaxies on the red sequence. 
They also find a low fraction of AGNs in the blue cloud but note that for blue galaxies their AGN sample is incomplete for $L_\mathrm{OIII}\lesssim10^{41}$ \ergs; at low luminosities the AGN emission lines are overwhelmed by the star formation precluding identification of the AGN. 
The blue galaxy sample is also generally at low stellar masses ($\log \mstel/\msun\lesssim10.5$).
Using our best-fit results for $p(\lamedd \giv \mstel,z)$ from Section \ref{sec:eddratio} and assuming a bolometric correction of $L_\mathrm{bol}\approx450L_\mathrm{OIII}$ \citep{Kauffmann09}, we predict $\lesssim1$\% of galaxies with $9.5\lesssim\log\mstel/\msun\lesssim10.5$ at $z<0.05$ to host AGN with $L_\mathrm{OIII}>10^{41}$.
Thus the results of S10 appear consistent with our findings, but the bias against the selection of AGNs in low-mass, star-forming galaxies in their work may have a significant effect.

\subsection{The role of AGNs in galaxy evolution}
\label{sec:role}

What do our results imply about the connection between AGN activity and galaxy evolution?
The fraction of galaxies hosting AGN rapidly decreases between $z\approx1$ and $z\approx0.2$ but follows the same distribution of Eddington ratios. 
We find strong evolution in the fraction of galaxies hosting AGN with redshift of the form $p(\lamedd \giv \mstel,z) \propto (1+z)^{3.5\pm0.5}$.
This evolution is very similar to the overall evolution in the star formation density, $\dot\rho_* \propto (1+z)^{3.5\pm0.2}$ for $0<z<1.2$ \citep{Rujopakarn10}.
This correspondence could be due to a relationship between the processes that fuel star formation and AGN: both could be fueled from the same gas supply,  or the AGN could be fueled indirectly via stellar mass loss from the stars within the galaxy \citep[e.g.,][]{Norman88,Ciotti07}.

We also find a weak association
between the prevalence of AGN activity and blue optical colors associated with current star formation activity.
That said, AGNs with a wide range of Eddington ratios are found in galaxies of all colors.
The \emph{shape} of the distribution of Eddington ratios is roughly constant as a function of color. 
The enhancement of AGN activity in galaxies with blue colors could indicate that the distribution of Eddington ratios is shifted to higher \Lamedd, on average.
This could imply that red galaxies contain AGNs at later stages of their lifetimes when the rate of accretion has reduced and the luminosities have faded. 
It has been proposed that feedback from an AGN can quench star formation in galaxies, resulting in the red colors \citep[e.g.,][]{Springel05,Hopkins06}.
However, the most rapidly accreting SMBHs (highest Eddington ratio sources) in our sample are \emph{not} exclusively in galaxies with current star formation (blue colors). 
Red colors and a lack of current star formation for a galaxy do not preclude substantial AGN accretion and significant SMBH growth.
Equally, a high fraction of blue galaxies contain weakly accreting (low Eddington ratio) AGNs. 

Thus, while AGN activity and star formation do appear to be correlated, we do not find evidence that AGNs play a direct role in shaping the global properties of their host galaxies or their evolution.
However, galaxies and their central AGNs could co-evolve in that the AGN fraction, related to the duty cycle of AGNs, may be regulated by whatever process drives gas to the centers of galaxies and fuels black hole growth. 
As discussed in Section \ref{sec:agntriggering} above, this process must result in a shift to lower characteristic accretion rates as cosmic time progresses. 
This does not prevent the triggering of AGN accretion in galaxies of any color but must lead to characteristically lower accretion rates 
in red galaxies \emph{or} a reduction in the triggering rate for red galaxies. 
As with the redshift evolution of the Eddington ratio distribution, we cannot distinguish between a shift to characteristically higher \Lamedd\ for blue galaxy colors or an overall increase in the density of blue galaxies hosting AGNs at a given \Lamedd.

We thus conclude that our results do not show evidence that feedback from moderate-luminosity AGNs plays a direct role in the evolution of galaxies, although star formation and AGN activity may have a common triggering mechanism.
We note that our results have no bearing on any potential role that may be played by luminous QSOs in the 
evolution of galaxies.

\subsection{Remaining issues and future prospects}
\label{sec:issues}

Despite our unprecedentedly large sample of galaxies and hard X-ray-selected AGNs with spectroscopic redshifts at $z\sim0.2-1$, our results are dominated by Poisson uncertainties.
This is due to the relatively small fraction of galaxies that host AGNs above a given X-ray luminosity ($\sim 1$\%$-10$\% but highly dependent on stellar mass and redshift) as well as the depths of our X-ray data which impede detection of lower luminosity AGNs.
We account for the Poissonian uncertainties in our analysis and are able to constrain functional forms that provide a good description of our data, although our discriminating power is limited by the source numbers.
Indeed, the results of Section \ref{sec:col} hint that the probability of hosting an AGN may have a complex color dependence that deviates from our global Eddington ratio dependence. However, due to our limited sample, we cannot split the data into many fine bins in redshift, stellar mass, X-ray luminosity and galaxy color at the same time to examine this behavior more closely.

We also note that our results are
valid for the population of moderately obscured, moderate-luminosity AGNs probed by our data.
This population appears to represent the majority of the AGN accretion density over our redshift range \citep{Ueda03,Aird10}, and thus our results can shed light on the dominant processes that are responsible for triggering and regulating AGN activity over this epoch. 
Our results should not be extrapolated to the very low or high Eddington ratio regime 
where different accretion and fueling processes may be relevant.
\moreamend{Furthermore, the lowest stellar mass galaxies in our sample are only included at the lower end of our redshift range due to
the redshift-dependent stellar mass cut, which is defined by the magnitude limits of our survey. 
This cut could introduce a bias in our results, although applying an additional cut of $\log \mstel/\msun>10.5$ at all redshifts does not significantly alter our results (but does substantially increase the errors due to the smaller sample size).
Conversely, higher stellar mass galaxies tend to be found at higher redshifts due to the larger volume surveyed, which could also introduce a bias. Both larger area surveys and deeper samples are required to address these issues.}

BLAGNs may dominate the population at the highest Eddington ratios, and thus their exclusion could also bias our derived Eddington ratio distribution. 
\amend{To investigate the impact of BLAGNs, we apply a completeness correction term in our maximum-likelihood fitting based on the fraction of BLAGNs within the total sample of X-ray detected PRIMUS sources at $0.2<z<1.0$. We apply this correction as a function of X-ray luminosity (over the $42<\log \lx<44$ range) but assume no redshift evolution and implicitly assume that the host stellar masses of the BLAGNs have the same distribution as the non--broad-line sample. Using this basic correction, we find a slightly flatter slope, $\gamma_E=0.62\pm0.04$, which is consistent with the slope derived in Section \ref{sec:eddratio} (within the errors) and indicates that the exclusion of BLAGNs does not introduce a significant bias.}
\amend{Nevertheless, this simple correction may not accurately account for this population.}
BLAGNs may correspond to a specific phase in the lifetimes of AGNs, be powered by different accretion modes, or be a distinct population that is triggered and fueled by very different processes compared to our sample.
\amend{Indeed, BLAGNs dominate the X-ray detected population at high luminosities ($\lx\gtrsim10^{44}$ \ergs), outside the range probed by our study and corresponding to the strong break seen in the XLF.
Such sources may dominate the AGN population at the highest Eddington ratios and constraints on their accretion properties are vital to track the shape of the Eddington ratio distribution and any cutoff around the Eddington limit. 
Estimates of the SMBH mass for such sources are possible using high-resolution spectroscopy and scaling relations between the luminosity, the width of broad lines and the SMBH mass \citep[e.g.,][]{Trump11}, which would allow estimates of the Eddington ratios and a more thorough investigation of the Eddington ratio distribution of BLAGNs. However, this is not possible with our PRIMUS spectra and is thus beyond the scope of this paper.}

Our hard X-ray selection may also bias our sample against the most absorbed AGNs ($\NH\gtrsim10^{23}$ cm$^{-2}$) 
and very heavily obscured, Compton-thick sources will 
be missed completely. 
While such sources may correspond to a substantial fraction of the AGN population \citep[e.g.,][]{Gilli07,Draper09}, it is unclear if they represent a distinct phase of AGN growth and thus how their co-evolution with galaxies may differ to the moderately obscured population.

Our work shows that selection effects can severely bias samples of AGNs selected using different methods due to the interplay between the Eddington ratio, the observed AGN luminosity, and the stellar mass and star formation properties of the galaxy.
However, by correcting for the various selection effects in our data, we are able to reveal the overall trends in the prevalence of AGNs and the dependence on redshift, stellar mass, X-ray luminosity or Eddington ratio.
Thus, our results can strongly constrain models of AGN triggering and the co-evolution of AGNs and galaxies.
Future work could push our studies to higher redshifts or compare our findings with relationships in the local universe.
Alternatively, increased area coverage and X-ray depth could increase our sample size and allow a more thorough analysis of the relationship between luminosity, absorption properties, \amend{bolometric corrections,} accretion rates, stellar masses, galaxy colors, star formation properties, and morphologies as a function of redshift over this key period of the universe, providing further insight into the physics of black hole growth and the relationship between black holes and their host galaxies.
However, any such future studies must carefully account for stellar-mass-dependent selection effects and the X-ray sensitivity to shed light on the true distribution of AGN accretion.


\section{Summary}
\label{sec:summary}

In this paper we study how the probability of a galaxy hosting an AGN depends on stellar mass and color, as well as revealing the underlying distribution of accretion rates probed by the X-ray luminosities and Eddington ratios.
We use an unprecedentedly large sample of galaxies with spectroscopic redshifts from PRIMUS, deep broad-band optical imaging, and X-ray data from \chandra\ or \xmm.

Our main conclusions are as follows:
\begin{itemize}

\item We find that for galaxies at $z=0.2-1$ the probability of hosting an
AGN as a function of X-ray luminosity, $p(\lx \giv \mstel,z)$, is a
power-law function of \Mstel\ with slope $\gamma_\mathcal{M}$ that is
independent of \Lx.
We also find that the distribution of X-ray luminosities at
fixed stellar mass is a power law with slope $\gamma_L$ that is
independent of stellar mass.
These results imply that while AGN
activity is enhanced in higher stellar mass sources, the
\emph{distribution} of accretion rates is independent of mass.

\item 
We find that the power-law slopes do not change
with redshift, but there is strong evolution of the overall
normalization. The probability of a galaxy hosting an AGN drops by a factor $\sim 6$ between $z\sim1$ and $z\sim0.2$.

\item We show that the
probability of a galaxy hosting an AGN is defined by a universal Eddington ratio
distribution, which is a power-law function of \Lamedd\ with slope
$\gamma_E=-0.65$ and is \emph{independent} of the \emph{stellar mass} of the host galaxy.

\item
Alternatively, these results can be interpreted in terms of \emph{specific accretion rate}---the rate of black hole growth relative to the \emph{stellar mass} of the host galaxy---rather than the true Eddington ratio. 
Thus, the distribution of \emph{specific accretion rates} is described by a universal power-law function with slope $-0.65$ for all stellar masses. 

\item 
The Eddington ratio distribution evolves strongly with redshift with the form $(1+z)^{3.5\pm0.5}$, similar to the rapid decrease in the total star formation rate density since $z\sim1$. 
However, the drop in the overall black hole accretion rate density is driven by this decrease in the probability of hosting an AGN for galaxies of \emph{all} stellar masses. 
We do not find a shift in the average stellar mass of a galaxy with an accreting black hole to lower masses or a shut down of AGN activity in the most massive galaxies as cosmic time progresses. 

\item We find that the probability of a galaxy hosting an AGN is mildly enhanced in galaxies with relatively \emph{blue} colors by a factor $\sim2$, once stellar-mass- and redshift-dependent selection effects are fully accounted for. 
However, AGN activity is taking place in galaxies of all colors.
The shape of the distribution of AGN accretion rates does not appear to significantly vary between the blue cloud, green valley and red sequence, and the distributions span the full range of Eddington ratios. 

\item AGNs are \emph{not} predominantly in red, massive galaxies. The previously observed increase in the AGN fraction with stellar mass is a selection effect that is driven by the \Lamedd\ distribution---a higher fraction of AGNs are detected in high stellar mass galaxies (with more massive black holes and generally redder colors) as sources accreting at low Eddington rates are intrinsically more luminous than in lower stellar mass galaxies.
\end{itemize}

Our results imply that the underlying physical processes that are responsible for triggering and fueling obscured AGN activity do not change between $z\sim1$ and $z\sim0.2$ but must decrease in frequency or shift to characteristically lower Eddington ratios. 
The same processes must take place in galaxies of all stellar masses and all colors, although activity appears to be enhanced in galaxies with current star formation.
While AGN activity and star formation appear to be globally correlated, we do not find evidence that AGN feedback plays a direct and important role in the quenching and regulation of star formation in galaxies.

\acknowledgements
We thank the referee for helpful comments that have improved this paper.
We also thank Aleks Diamond-Stanic for helpful discussions and comments on this paper.
We acknowledge 
Rebecca Bernstein, Adam Bolton, Douglas Finkbeiner, David W. Hogg, Timothy McKay, Alexander J. Mendez, Sam Roweis, and Wiphu Rujopakarn
for their contributions to the PRIMUS
project. We thank the CFHTLS, COSMOS, DLS, and 
SWIRE teams for their public data releases and/or access to early releases.  
This paper includes data gathered with the 6.5m Magellan 
Telescopes located at Las Campanas Observatory, Chile.
We thank the support staff at LCO 
for their help during our observations, and we acknowledge
the use of community access through NOAO observing time.
Some of the data used for this project are from the CFHTLS public data release,
which includes observations obtained with MegaPrime/MegaCam, a joint project 
of CFHT and CEA/DAPNIA, at the Canada--France--Hawaii Telescope (CFHT) which is 
operated by the National Research Council (NRC) of Canada, the Institut 
National des Science de l'Univers of the Centre National de la Recherche 
Scientifique (CNRS) of France, and the University of Hawaii. This work is 
based in part on data products produced at TERAPIX and the Canadian Astronomy 
Data Centre as part of the Canada-France-Hawaii Telescope Legacy Survey, a 
collaborative project of NRC and CNRS.
We also thank those who have built and operate the \textit{Chandra} and \textit{XMM-Newton} X-ray observatories.
Funding for PRIMUS has been provided
by NSF grants AST-0607701, 0908246, 0908442,
and 0908354, and NASA grant 08-ADP08-0019.
R.J.C. is supported by NASA through Hubble Fellowship grant 
HF-01217 awarded by the Space Telescope Science Institute, 
which is operated by the Association of Universities for
Research in Astronomy, Inc., for NASA, under contract NAS 5-26555.

\bibliographystyle{apj}

\end{document}